\documentclass[fleqn,usenatbib]{mnras}

\usepackage{mathptmx}

\usepackage[T1]{fontenc}

\DeclareRobustCommand{\VAN}[3]{#2}
\let\VANthebibliography\thebibliography
\def\thebibliography{\DeclareRobustCommand{\VAN}[3]{##3}\VANthebibliography}

\usepackage{graphicx}	
\usepackage{amsmath}	
\usepackage{amssymb}	
\usepackage{nicefrac}
\usepackage{enumitem}
\setlist{itemsep=1mm}

\newcommand{\MYhref}[3][blue]{\href{#2}{\color{#1}{#3}}}%
\newcommand{\MYhbar}{{\raisebox{-0.05ex}{$\mathchar '26$}\mkern -7muh}}%
\newcommand{\MYsmhbar}{{\raisebox{-0.35ex}{$\mathchar '26$}\mkern -8muh}}%

\mathchardef\mhyphen="2D

\title[A precise photometric ratio II]{A precise photometric ratio via laser excitation of the sodium layer -- II. \\
                                      {\fontsize{13.2}{15.8}\selectfont Two-photon excitation using lasers detuned from 589.16~nm and 819.71~nm resonances}}

\author[J. E. Albert et al.]{Justin E. Albert \href{https://orcid.org/0000-0003-0253-2505}{\includegraphics[scale=0.55]{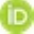}},$^{1}$\thanks{E-mail: \href{mailto:jalbert@uvic.ca}{jalbert@uvic.ca} (JEA)}
Dmitry Budker \href{https://orcid.org/0000-0002-7356-4814}{\includegraphics[scale=0.55]{orcid}},$^{2,4,5}$
Kelly Chance \href{https://orcid.org/0000-0002-7339-7577}{\includegraphics[scale=0.55]{orcid}},$^{3}$
Iouli E. Gordon \href{https://orcid.org/0000-0003-4763-2841}{\includegraphics[scale=0.55]{orcid}},$^{3}$\newauthor
Felipe Pedreros Bustos \href{https://orcid.org/0000-0001-7664-1590}{\includegraphics[scale=0.55]{orcid}},$^{2,6}$
Maxim Pospelov,$^{1,7,\ddagger}$
Simon M. Rochester \href{https://orcid.org/0000-0001-5202-5718}{\includegraphics[scale=0.55]{orcid}}\hspace*{0.35mm}$^{4}$
and H. R. Sadeghpour \href{https://orcid.org/0000-0001-5707-8675}{\includegraphics[scale=0.55]{orcid}}\hspace*{0.35mm}$^{3}$
\\
$^{1}$Department of Physics and Astronomy, University of Victoria, Victoria, British Columbia V8W 3P6, Canada\\
$^{2}$Helmholtz Institute, Johannes Gutenberg-Universit\"at Mainz, 55099 Mainz, Germany\\
$^{3}$Harvard-Smithsonian Center for Astrophysics, Cambridge, Massachusetts 02138, USA\\
$^{4}$Rochester Scientific LLC, El Cerrito, California 94530, USA\\
$^{5}$Department of Physics, University of California, Berkeley, California 94720-7300, USA\\
$^{6}$Laboratoire d'Astrophysique de Marseille (LAM), Universit\'e d'Aix-Marseille \& CNRS, F-13388 Marseille, France\\
$^{7}$Perimeter Institute of Theoretical Physics, Waterloo, Ontario N2L 2Y5, Canada\\
$^{\ddagger}$Now at School of Physics and Astronomy, University of Minnesota, Minneapolis, Minnesota 55455, USA
\vspace*{-3mm}
}

\date{Accepted 2021 May 31. Received 2021 May 20; in original form 2021 January 18}

\pubyear{2021}

\begin{document}
\jname{\MYhref[magenta]{https://doi.org/10.1093/mnras/stab1619}{\textbf{MNRAS}}}
\volume{508}
\label{firstpage}
\pagerange{\MYhref[blue]{https://doi.org/10.1093/mnras/stab1619}{4412--4428}}
\maketitle

\begin{abstract}
This paper is the second in a pair of papers on the topic of the generation of a two-colour artificial star
(which we term a ``laser photometric ratio star,'' or LPRS) of de-excitation light from neutral sodium atoms in the mesosphere,
for use in precision telescopic measurements in astronomy and atmospheric physics, and more specifically for the
calibration of measurements of dark energy using type~Ia supernovae.  The two techniques respectively described in both
this and the previous paper would each generate an LPRS with a precisely 1:1 ratio of yellow~(589/590~nm) photons to
near-infrared~(819/820~nm) photons produced in the mesosphere. Both techniques would provide novel mechanisms for establishing
a spectrophotometric calibration ratio of unprecedented precision, from above most of Earth's atmosphere,
for upcoming telescopic observations across astronomy and atmospheric physics;
thus greatly improving the performance of upcoming measurements of dark energy parameters using type~Ia supernovae.
The technique described in this paper has the
advantage of producing a much brighter (specifically, brighter by approximately a factor of $10^3$) LPRS, using
lower-power ($\le$30~W average power) lasers, than the technique using a
single 500~W average power laser described in the first paper of this pair.  However, the technique described here
would require polarization filters to be installed into the telescope camera in order to sufficiently remove laser
atmospheric Rayleigh backscatter from telescope images, whereas the technique described in the first paper
would only require more typical wavelength filters in order to sufficiently remove laser Rayleigh backscatter.
\end{abstract}

\begin{keywords}
techniques:photometric -- methods:observational -- telescopes -- instrumentation:miscellaneous -- dark energy
\end{keywords}



\defcitealias{Alb20a}{Paper~I}

\section{Introduction}

The motivations for the generation of a laser photometric ratio star (LPRS) are detailed in the first paper in this pair of  
papers~\citep[][hereafter referred to as \citetalias{Alb20a}]{Alb20a}.  To briefly review: measurements of dark energy 
using type~Ia supernovae (SNeIa) are limited by systematic uncertainty on astronomical magnitude as a function of colour,
within the optical spectrum~\citep{Jones18,Betoule14,WoodVasey07}; the generation of an LPRS above an SNeIa-survey telescope   
could provide a precise calibration source to effectively eliminate this dominant uncertainty.  Additionally, other
types of astronomical measurements (besides SNeIa cosmology) could greatly benefit from reduction in relative photometric
uncertainty~\citep{Connor17,Kirk15} via LPRS-based calibration.  In this paper, we detail a technique for the generation of
an LPRS that utilises two lasers, at optical frequencies that are, respectively, approximately 3.9~GHz below and above
neutral sodium atomic resonances that occur at wavelengths of 589.16~nm and 819.71~nm.\footnote{In this paper, wavelengths
are given in vacuum, typically to either the nearest nanometre, or nearest hundredth of a nanometre.  Wavelengths for sodium
are as provided by~\citet{Kel08}.}  As we will show,
this technique for LPRS generation will result in a significantly brighter LPRS with an apparent magnitude of
approximately 12 (as compared with magnitude 20), using much lower-power lasers, than the technique described in~\citetalias{Alb20a}.

\section{Laser Photometric Ratio Star (LPRS) Using Two Lasers Detuned From N\lowercase{a}\,{\small I} Atomic Resonances}

\begin{figure}
\begin{center}
\includegraphics[scale=1.1]{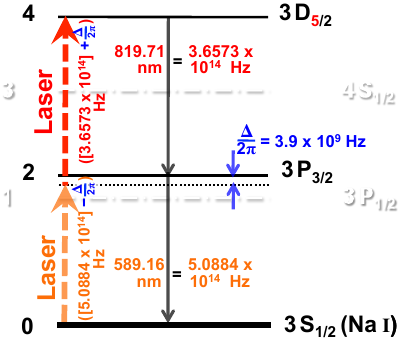}
\end{center}
\caption{Atomic level diagram (not to scale) for neutral sodium atoms in the two-laser LPRS technique described in this paper.
The allowed transitions and levels (i.e., levels 0, 2, and 4 in the diagram) are in solid black, whereas
a dotted black line represents the off-resonant energy corresponding to the frequency of
the first laser.  The 5 in the allowed $3\,$D$_{5/2}$ level is
red to distinguish this atomic state from the $3\,$D$_{3/2}$ \ion{Na}{i} state
that is excited by the 343~nm laser in the technique described in~\citetalias{Alb20a}.
The two ``ghost'' levels (1 and 3) below the $3\,$D$_{5/2}$ excited state, that are both inaccessible from states in
the diagrams that are excited by the lasers, are shown in shadowed gray text and dash-dotted lines.
As is shown, this LPRS technique results
in a ``fully-mandated cascade'' from the 819.71~nm de-excitation to the 589.16~nm de-excitation, resulting in a
mandated 1:1 ratio between those produced photons.
}
\label{fig:LPRSPrimeDiagram}
\end{figure}

Figure~\ref{fig:LPRSPrimeDiagram} outlines how neutral sodium atoms (\ion{Na}{i}) in the ground state can undergo photoexcitation
to the $3\,$D$_{5/2}$ state via two-photon absorption using pulses from two lasers, with one laser adjusted to a wavelength
slightly greater than the \ion{Na}{i} resonance at 589.16~nm and the second laser adjusted to a wavelength that is correspondingly
smaller than the \ion{Na}{i} resonance at 819.71~nm.  The value of the selected ``detuning parameter'' $\Delta$ can be
modified: as we show quantitatively below, smaller values of $\Delta$ would result in
a brighter LPRS.  However, if $\frac{\Delta}{2\pi}$ were chosen smaller than approximately 3.3~GHz, then at least
1\% of the observed photons from the LPRS \ion{Na}{i} de-excitation would result from direct single-photon excitation
to the $3\,$P$_{3/2}$ state rather than from two-photon excitation to the $3\,$D$_{5/2}$ state, which would serve to spoil the
precise 1:1 photometric ratio characteristic of the resulting LPRS.  We have, thus, selected a value of
$\frac{\Delta}{2\pi} = 3.9$~GHz, so that a smaller fraction than approximately 2 out of $10^5$ observed photons from the \ion{Na}{i}
de-excitation in the LPRS will result from single-photon excitation to the $3\,$P$_{3/2}$ state, rather than from the
intended two-photon excitation to the $3\,$D$_{5/2}$ state.

\subsection{Backgrounds from Rayleigh scattering and from virtual 3$\,\text{P}^{*}_{3/2}$ production, and the need for STIRAP}
\label{subsec:BkgsNeedForSTIRAP}

Before providing a formula for the excitation rate of the signal process shown above in Fig.~\ref{fig:LPRSPrimeDiagram} as a function of the
detuning parameter and of the two laser intensities, we critically note that, if other laser parameters besides the detuning and intensity
(i.e.~pulse timings, shapes, and polarizations)
were not chosen carefully, there would be two forms of dominant backgrounds to the process: near-180\degr~atmospheric
Rayleigh back-scattering of light from the lasers; and the de-excitation of virtual \ion{Na}{i}
excitations, which we denote as $3\,$P$^{*}_{3/2}$ (located near
in frequency to the non-virtual $3\,$P$_{3/2}$ state) that would be excited in the sodium layer during pulses by the 589.16~nm laser.
We discuss backgrounds further, and calculate their expected rates, in Section~\ref{sec:EstBkgd} of this paper, but in order to first
begin to consider the outline of an optimal system to reject as much
background light, but generate and accept as many signal photons, as is possible, we must first consider some
basic choices in pulse timings, shapes, and polarizations from the two lasers.

The background from near-180\degr~Rayleigh back-scattering of 589.16~nm and 819.71~nm light from the two lasers will result in a
large flux of photons of those two wavelengths into the telescope aperture.  Note that this is a similar issue to the Rayleigh scattering background
encountered in~\citetalias{Alb20a}; however, the problem is compounded in the present situation by the fact that
here these are nearly the same two wavelengths as the signal photons from the LPRS itself.  Thus, in this case one cannot just improve the
out-of-band rejection of the telescope filters (as one could in that previous situation where a single 342.78~nm laser was used).  The main handle in
the present situation for the rejection of Rayleigh back-scattered background photons will, thus, need to be polarization, rather than wavelength.
I.e., if the 589.16~nm laser has a given polarization, then the telescope $r$ filter
(that accepts 589.16~nm light) must \textit{block} that particular polarization, and only accept the orthogonal polarization, thereby blocking a large majority
of Rayleigh back-scattered light from that laser.  Similarly, if the 819.71~nm laser has a given polarization, then the telescope $i$ and $z$ filters
(that accept 819.71~nm light) must block \textit{that} particular polarization, and only accept the orthogonal polarization.
These laser and filter polarizations could, of course, be linear or circular.  For a consistent definition of our linear polarization orientations,
we define the $\hat{x}$ and $\hat{y}$ orientations to respectively be
in the east-west and in the north-south geographical directions for the sodium fluorescence light propagating toward the nadir that enters a zenith-pointing telescope
and the $+\hat{z}$ direction to always be in the direction of the propagation of the relevant light that is under consideration within the given context --- despite
the fact that this Cartesian axis frame of course changes with respect to Earth and the direction of gravity, depending on the light that is under
consideration.  We shall choose an $\hat{x}$ linear polarization for the light output of the 589.16~nm laser
and a $\hat{y}$ linear polarization for the light output of the 819.71~nm laser;
and thus the modified telescope $r$ filter must reject $\hat{x}$-polarized light and only accept $\hat{y}$-polarized light,
and the modified telescope $i$ and $z$ filters must both reject $\hat{y}$-polarized light and only accept $\hat{x}$-polarized light.

If care were not
additionally taken with the laser pulse shapes, and with the relative timing of the pulses of the 589.16~nm and 819.71~nm lasers (or, for that matter, if either one of the lasers
were continuous-wave, rather than pulsed), a large rate of virtual $3\,$P$^{*}_{3/2}$ \ion{Na}{i} excitations would occur during the 589.16~nm laser pulses, followed by a decay
back down to the ground state and emission of background 589.16~nm light.  Specifically, during each 589.16~nm laser pulse, this
laser-induced background de-excitation rate (the number of de-excitation events from virtual $3\,$P$^{*}_{3/2}$ excitations
per \ion{Na}{i} atom per unit time) would be approximately given by:
\begin{eqnarray}
\label{Eq:DeExRateBkgdNoSTIRAP}
W^{\rm Na\,{\scriptscriptstyle I}\,(bkgd.\;de\mhyphen excitation\;during\;pulse,\;if\;no\;STIRAP)}_{(3\,{\rm P}^{*}_{3/2}) \; \to \; (3\,{\rm S}_{1/2}) + \gamma_{\rm 589\,nm}} \quad \approx \qquad & & \nonumber\\
    \frac{1}{4 \MYhbar^{2}} \frac{g_k}{g_i}
    \left[ \frac{d^2_{ik} \mathcal{E}^2_1\:\!\!(x,y,t)}{\Delta^2} \right] \Gamma_k, \; \quad \qquad & &
\end{eqnarray}   
where the detuning parameter $\Delta$ was defined previously; states $\{i,k\} \equiv \{3\,$S$_{1/2},\:3\,$P$_{3/2}\}$;
the level degeneracy ratio $\frac{g_k}{g_i} = \frac{4}{2} = 2$;
the natural linewidth $\Gamma_k$ of state $k$ is $6.16 \times 10^7$~s$^{-1}$;
the dipole moment $d_{ik} = \sqrt{\frac{3\epsilon_0 h c^3 A_{ki}}{2\omega^3_{ki\phantom{f}}}} = 2.11 \times 10^{-29}$~coulomb-metres
(where the Einstein $A$ coefficient $A_{ki} = 6.16 \times 10^7$~s$^{-1}$);
and the average electric field strength due to the the 589.16~nm laser beam as a function of the
transverse distances $(x,y)$ from the beam centreline\footnote{\label{foo2}The small divergence of the laser beams implies that the average electric field
strength within the sodium layer will also depend on [in addition to depending on $(x,y,t)$]
the path length that the beam has taken through the atmosphere,
however we average over that effect in our flux calculations in Sections~\ref{sec:EstFlux} and~\ref{sec:EstBkgd}.}
and the time $t$ after the passage of the midpoint between a given pair of pulses from the two lasers,
$\mathcal{E}_1\:\!\!(x,y,t)$, is provided in V$\!/$m within the altitude range of the sodium layer~\citep{Bud08}.
This \textit{nominal} de-excitation rate for virtual $3\,$P$^{*}_{3/2}$ excitations,
and resulting background, is larger than our signal excitation rate.
However, the use of the laser optical technique known as STIRAP [STImulated Raman Adiabatic Passage, \citet{Gau90}] will allow one to
nearly completely avoid the production of virtual $3\,$P$^{*}_{3/2}$ \ion{Na}{i}
excitations and thus of their associated background de-excitation light.

The STIRAP technique [reviewed by \citet{Vit17}], as specifically considered here, involves pulsing the 819.71~nm laser and then the 589.16~nm laser in
succession, such that \ion{Na}{i} atoms in the $3\,$S$_{1/2}$ ground state are adiabatically transferred by two photons up to the $3\,$D$_{5/2}$ state, without
ever landing in the intermediate virtual $3\,$P$^{*}_{3/2}$ excitation.  In practice, with (for example) pulses from the 819.71~nm laser and the 589.16~nm laser
that are each temporally Gaussian-distributed with $\sigma_t^{\rm 819.71\;nm} = \sigma_t^{\rm 589.16\;nm} = 1$~ns, this would involve delaying the Gaussian
peak of each 589.16~nm laser pulse to approximately $\sqrt{2}$~ns \textit{after} the peak of each 819.71~nm laser pulse.  (We importantly note that this
ordering is, at least at first glance,
counter-intuitive: the laser at the frequency that is near to the excitation frequency from the ground state to the intermediate state should
peak in time \textit{following} the peak of the other laser.)  The STIRAP technique has been experimentally demonstrated in many laboratory results since
1990 to have adiabatic transfer efficiencies of nearly 100\%, and also to be robust to small experimental variations in laser parameters~\citep{Ber15}.
STIRAP has been performed at gas pressures up to atmospheric pressure at sea level [on sodium atoms within an argon buffer gas,~\citet{Joh10}],
however STIRAP has not yet been demonstrated in the open atmosphere.

\subsection{Effective signal excitation rate}

The condition that must be satisfied in order for STIRAP excitations of ground state \ion{Na}{i} atoms to the $3\,$D$_{5/2}$ excited state   
to occur, due to a given single pair of pulses from the two lasers,
at locations that are a transverse distance$^{\ref{foo2}}$ $(x,y)$ from the centreline of the laser beams, is
\begin{equation}
\label{Eq:STIRAPGlobalAdiabaticCond}
\mathcal{A}(x,y) \: \equiv \:
\frac{1}{\MYhbar} \!\!
\int\limits^{\infty}_{-\infty} \!\!\!\left( \! \sqrt{d^2_{ik}\mathcal{E}^2_1\:\!\!(x,y,t) \; + \; d^2_{kf}\mathcal{E}^2_2\:\!\!(x,y,t)}\; \right) dt
\;\;\; \gg \;\;\; \frac{\pi}{2}\,,
\end{equation}
where $d_{ik}$, $\mathcal{E}_1$, and states $i$ and $k$ are as defined previously; state $f \equiv 3\,$D$_{5/2}$;
$d_{kf} = \sqrt{\frac{3\epsilon_0 h c^3 A_{fk}}{2\omega^3_{fk}}} = 3.17 \times 10^{-29}$~coulomb-metres
(where the Einstein $A$ coefficient $A_{fk} = 5.14 \times 10^7$~s$^{-1}$);
and the average electric field strength $\mathcal{E}_2$ due to the the 819.71~nm laser beam as a function of $(x,y)$
and of the time $t$ after the passage of the midpoint between a given pair of pulses from the two lasers
is provided in V$\!/$m within the altitude range of the sodium layer~\citep{Vit17}.
The quantity $\mathcal{A}$,
a function of $x$ and $y$, is known as the ``pulse area,'' although one should note that
the values of $\mathcal{A}(x,y)$ itself are dimensionless, rather than having dimensions of area.
The condition $\mathcal{A} \gtrsim 10$ is generally sufficient for efficient STIRAP population transfer in laboratory measurements.

The condition in equation~(\ref{Eq:STIRAPGlobalAdiabaticCond}) will result in a column of sodium atoms through the sodium layer being
excited to the $3\,$D$_{5/2}$ state
each time that a pair of pulses from the lasers passes by; and, if the time between successive pairs of laser pulses is long compared
with the total decay time (i.e., compared with about 40~ns) from $3\,$D$_{5/2}$ back down to the $3\,$S$_{1/2}$ ground state, the resulting two quantites
that are relevant for the effective rate of emission of 819.71~nm and 589.16~nm photons from the sodium atoms in the mesosphere, that are
governed by the parameters of the two utilised lasers, will be: 1) The cross-sectional area of the column through the sodium layer for which the
condition in equation~(\ref{Eq:STIRAPGlobalAdiabaticCond}) holds true, and 2) The time interval between successive pairs of laser pulses.
We will determine those two quantities, and we will thus estimate the effective rate of emission of 819.71~nm and 589.16~nm photons
from sodium in the mesosphere, in Section~\ref{sec:EstFlux}.

\subsubsection{Doppler shift detuning and signal excitation fraction\label{subsubsec:DopDet}} However, the important effect on STIRAP signal excitation efficiency
from Doppler shifts due to the velocity distribution of the sodium atoms in the mesosphere, and from the linewidths of the
lasers, must additionally be considered.  Determination of the results of these effects on STIRAP efficiency is
non-trivial, and multiple methods by various authors [reviewed in~\citet{Vit17}] have been developed for analytic and numerical approximation of the results.
If, for a given pair of photons in the mesosphere that are respectively from the 589~nm and the 820~nm lasers, we consider the
angular frequency detunings from the peak of the 589.16~nm and the 819.71~nm resonances to respectively be $\Delta_{\rm 589\,nm}$ and
$\Delta_{\rm 820\,nm}$, and a given \ion{Na}{i} atom in the mesosphere to have a velocity vector $\vec{v}$, then the resulting velocity-dependent
angular frequency detuning $\delta_{\rm eff}$ from the peak of the two-photon STIRAP resonance will be given by
\begin{equation}
\qquad \quad \; |\delta_{\rm eff}| = |\Delta_{\rm 589\,nm} - \Delta_{\rm 820\,nm} + (\vec{k}_{\rm 589\,nm} + \vec{k}_{\rm 820\,nm})\cdot\vec{v}|,
\end{equation}
where $|\vec{k}_{\rm 589\,nm}| = \frac{2\pi}{\lambda_{\rm 589\,nm}}$ and $|\vec{k}_{\rm 820\,nm}| = \frac{2\pi}{\lambda_{\rm 820\,nm}}$ are
the wavevectors of the two respective photons.  Thus, in the case of perfectly null two-photon detuning
(i.e., when $\Delta_{\rm 589\,nm} = \Delta_{\rm 820\,nm}$) if, for example, the two photons happen to reside at the peaks of the two laser lines, we will have
\begin{equation}
\qquad \qquad \qquad |\delta^{\rm Doppler}_{\rm eff}| = (|\vec{k}_{\rm 589\,nm}| + |\vec{k}_{\rm 820\,nm}|)|v_{\hat{z}}|,
\end{equation}
where $v_{\hat{z}}$ is the component of the \ion{Na}{i} atomic velocity $\vec{v}$ along the laser propagation direction $\hat{z}$.
The STIRAP process is highly sensitive to this velocity-dependent detuning $\delta_{\rm eff}$: \ion{Na}{i} atoms in the mesosphere
that are within the spatial region given by equation~(\ref{Eq:STIRAPGlobalAdiabaticCond}) and which happen to have a small value
of $\delta_{\rm eff}$ with respect to pairs of photons from the two lasers will successfully be excited to the $3\,$D$_{5/2}$ state via STIRAP,
whereas the \ion{Na}{i} atoms which do not happen to reside within both this spatial region and
this narrow range of velocity along the $\hat{z}$ component direction will fail to
be excited via the STIRAP process.  The root mean square of the $v_{\hat{z}}$ distribution
$v^{\rm RMS}_{\hat{z}} = \sqrt{\frac{k_B T_{\rm 100\,km}}{M_{\rm Na}}}$ where $T_{\rm 100\,km} \approx 200$~K,
and thus $v^{\rm RMS}_{\hat{z}} \approx 270$~m/s.  Thus, $\delta^{\rm RMS,\,Doppler}_{\rm eff} \approx 4.93 \times 10^9$~s$^{-1}$.
The total effective root mean square detuning $\delta^{\rm RMS}_{\rm eff}$ will equal the sum in quadrature
of $\delta^{\rm RMS,\,Doppler}_{\rm eff}$ and the root mean square linewidths of the lasers (which we will take to each be 1~GHz =
$6.28 \times 10^9$~s$^{-1}$), and thus $\delta^{\rm RMS}_{\rm eff} \approx 1.02 \times 10^{10}$~s$^{-1}$.

The associated width of the two-photon resonance $\delta_{\nicefrac{1}{2}}$ that corresponds to a STIRAP transition probability of 50\% is
estimated in~\citet{Dan94} to be given, for Gaussian laser pulses from the two lasers which happen to have equal peak Rabi frequencies, by
\begin{equation}
\label{Eq:STIRAP2GamResonanceWidth}
\qquad \qquad \qquad \qquad \quad \; \delta_{\nicefrac{1}{2}}\sigma_t \:\: = \:\: A(\Omega_0 \sigma_t)^n,
\end{equation}
where $\sigma_t$ is the temporal length of the pulses from each of the two lasers; and
the peak Rabi frequency $\Omega_0 \equiv \frac{d_{\alpha\beta}\mathcal{E}_0}{2\MYsmhbar}$ where $d_{\alpha\beta}$ is the
dipole moment between the initial and intermediate state or the intermediate and final state, and $\mathcal{E}_0$ is the peak
electric field within the mesosphere of the first laser or the second laser.
$A$ and $n$ are both dimensionless, $\mathcal{O}(1)$ constants that are tabulated in~\citet{Dan94} and happen to both
be approximately equal to 0.9 for the ranges of experimental parameters that we will consider in this paper.
We will find in Section~\ref{sec:EstFlux} that this two-photon resonance width $\delta_{\nicefrac{1}{2}} \ll \delta^{\rm RMS}_{\rm eff}$,
and specifically that the root mean square detuning $\delta^{\rm RMS}_{\rm eff} \approx 1.02 \times 10^{10}$~s$^{-1}$ tends to
be around an order of magnitude larger than $\delta_{\nicefrac{1}{2}}$ for the ranges of experimental parameters that we will be
considering.

This implies that this two-photon resonance peak (that is parameterized by $\delta_{\nicefrac{1}{2}}$) effectively carves a narrow region
of the velocity distribution from the wider \ion{Na}{i} detuning spectrum (parametrized by $\delta^{\rm RMS}_{\rm eff}$) for STIRAP
excitation, and leaves the rest of the sodium atoms within the laser beam column in the mesosphere unexcited.  The fraction
$f_{\rm STIRAP}$ of the sodium atoms within the column in the mesosphere
that happen to be within that narrow region of the velocity distribution will, of course, be proportional to
$\frac{\delta_{\nicefrac{1}{2}}}{\delta^{\rm RMS}_{\rm eff}}$.  To find the constant of proportionality $C$ such that
$f_{\rm STIRAP} = C\frac{\delta_{\nicefrac{1}{2}}}{\delta^{\rm RMS}_{\rm eff}}$, we note that if both
$\delta_{\nicefrac{1}{2}}$ and~$\delta^{\rm RMS}_{\rm eff}$ happen to parametrize Gaussian distributions, then
$\delta^{\rm RMS}_{\rm eff}$ will be equal to $1\sigma$ of its Gaussian distribution; whereas $\delta_{\nicefrac{1}{2}}$ is a half-width at
half-maximum, i.e.~$\sigma\sqrt{2\ln2}$ of its Gaussian distribution.  The value of an integral within a region of width
$w$ that is carved from the centre of a broad normal distribution with standard deviation $\sigma \gg w$ is, of course,
$\frac{w}{\sigma\sqrt{2\pi}}$.  Thus, when the LPRS system happens to be on-centre of the detuning distribution,
i.e.~at an optimal two-photon laser tuning, the  
constant of proportionality $C = \frac{1}{\sqrt{2\pi}} \times \frac{1}{\sqrt{2\ln2}}$, and thus
\begin{equation}
\label{Eq:STIRAPExcitationFraction}
\qquad \qquad \qquad \quad \; f_{\rm STIRAP} \; = \; \frac{\delta_{\nicefrac{1}{2}}}{2\delta^{\rm RMS}_{\rm eff}\sqrt{\pi\ln2}}
\end{equation}
under the approximation of Gaussian detuning and two-photon resonance spectra.

Note that various methods such as laser pulse chirping~\citep{Ped20}, or synchronization of pulses (or their polarization) with the Larmor precession of
\ion{Na}{i} in the geomagnetic field~\citep{Kan14, Fan16, Ped18}, could potentially be used to, in effect, increase the above ratio, thus potentially increasing
LPRS brightness.  We do not, however, assume the implementation of such possible LPRS brightness-increasing techniques in this paper.  (We will find in
Sections~\ref{sec:EstFlux}~--~\ref{sec:EstImp} that this LPRS will be sufficiently bright that such enhancements should likely not be necessary.)\vspace*{-4mm}

\subsubsection{Signal photon polarization} Another parameter that is relevant, not for the effective rate of STIRAP
excitation and resulting
emission of the 819.71~nm and 589.16~nm photons, but rather for the efficiency
of their detection by the telescope camera approximately 95~km below, is the polarization of those emitted photons.  As mentioned in the previous
subsection, the light from the 589.16~nm and the 819.71~nm lasers will be linearly polarized in the
$\hat{x}$ and $\hat{y}$ directions respectively (and the telescope filters will reject
$\hat{x}$-polarized 589.16~nm light and $\hat{y}$-polarized 819.71~nm light, and only accept
$\hat{y}$-polarized 589.16~nm light and $\hat{x}$-polarized 819.71~nm light).
Due to the fact that the Doppler-broadened linewidth $\frac{\Gamma_{D_f}}{2\pi} \approx 1.5$~GHz is
of the same order as, or is sigificantly greater than, the separations of the hyperfine levels within each of the \ion{Na}{i} states $i$, $k$, and $f$
that are defined above (where the Doppler linewidth is of the same order for the case of the hyperfine separation within the ground state $i$,
and is significantly greater for the cases of the separations within the excited states $k$ and $f$), we may safely average over the
individual hyperfine levels within each state, when calculating the averaged emitted polarizations of the photons from the sodium layer.
Both the 819.71~nm and the 589.16~nm signal photons that are emitted
will be approximately equally polarized in the $\hat{x}$ and $\hat{y}$ directions (i.e., will be nearly unpolarized).
A simplified simulation, using the {\tt Atomic Density Matrix} software package~\citep{ADM}, of the LPRS system that we describe in the present
paper shows that there will be a small excess (approximately 0.3\%) of 819.71~nm signal photons emitted from the sodium layer that will be
polarized in the $\hat{y}$ direction; however this small excess (and the resulting small deficit in signal photons that successfully pass
through the the polarized telescope filter) do not significantly affect any of the results of this paper.
The necessary addition of linear polarization filters within the telescope's optical filters, in order to filter out what would be an
otherwise-dominant atmospheric Rayleigh scattering background does, however, cause the very important
loss of just over 50\% of signal photons (whether from the LPRS within the upper atmosphere, or from astronomical sources).  We include this
substantial effect when calculating expected numbers of detected signal photoelectrons in Section~\ref{sec:EstFlux} of this paper, as well
as in subsequent analysis.\vspace*{-4mm}

\subsubsection{Signal excitation to ${\rm 3}\,{\rm D}_{3/2}$, rather than to ${\rm 3}\,{\rm D}_{5/2}$} A small fraction
of the sodium atom excitations from the pairs of laser pulses will be to the $3\,$D$_{3/2}$ state, rather than to the $3\,$D$_{5/2}$ state.
This is due to the fact that the
the $3\,$D$_{3/2}$ excitation of \ion{Na}{i} resides only 1.5~GHz above the intended $3\,$D$_{5/2}$
excitation and, thus, there will be overlap between these two \ion{Na}{i} excitation frequency distributions due to
the aforementioned Doppler broadening.  Fortunately,
the presence of these $3\,$D$_{3/2}$ excitations in addition to the intended $3\,$D$_{5/2}$ excitations will not affect
the 1:1 ratio of yellow~(589/590~nm) photons to near-infrared~(819/820~nm) photons from the resulting LPRS, since both the
$3\,$D$_{3/2}$ and $3\,$D$_{5/2}$
excitations will produce mandated cascades of photons of those two wavelength ranges (the former as shown 
in Fig.~\href{https://arxiv.org/pdf/2001.10958.pdf#fig1}{1(c)} 
in~\citetalias{Alb20a}, and the latter as shown in Fig.~\ref{fig:LPRSPrimeDiagram} above in this paper).
These $3\,$D$_{3/2}$ excitations will be a small subset of the sodium atoms within the column described by equation~(\ref{Eq:STIRAPGlobalAdiabaticCond}),
rather than being within an additional region in excess to it, and thus it is only necessary to consider the total number of sodium
atoms in the region described by equation~(\ref{Eq:STIRAPGlobalAdiabaticCond}), rather than any additional conditions for $3\,$D$_{3/2}$ excitations.

\subsection{Excitation rates of possible significant laser-induced backgrounds other than Rayleigh scattering or
virtual 3$\,$P$^{*}_{3/2}$ excitation production}

In addition to the backgrounds described in subsection~\ref{subsec:BkgsNeedForSTIRAP},
two other forms of potentially-significant laser-induced background excitation rates, consisting of single-photon and three-photon
transition events to the non-virtual $3\,$P$_{3/2}$ state, per ground-state \ion{Na}{i} atom per unit time, will respectively be given by:
\begin{eqnarray}
\label{Eq:ExRateBkgd1}
W^{\rm Na\,{\scriptscriptstyle I}\,(other\;bkgd.\;excitation\;\#1)}_{(3\,{\rm S}_{1/2}) + \gamma_{\rm 589\,nm} \; \to \; (3\,{\rm P}_{3/2})} \:\, & \approx & \!\!\!\!\!\!\!\!
   \frac{\sqrt{\pi}}{4\MYhbar^{2}\Gamma_{D_k}\!\!} \frac{g_k}{g_i} \!\!
   \left( \!\! d^2_{ik} \mathcal{E}^2_1 \!\! \right) \!\! e^{-\Delta^2\!\!/\Gamma^2_{D_k}}\!, \:\: {\rm and}\quad\quad \\
\label{Eq:ExRateBkgd2}
W^{\rm Na\,{\scriptscriptstyle I}\,(other\;bkgd.\;excitation\;\#2)}_{(3\,{\rm S}_{1/2}) + 2\gamma_{\rm 589\,nm} \; \to \; (3\,{\rm P}_{3/2}) + \gamma_{\rm 589\,nm}} 
\!\!\!\!\!\!\!\!\!\!\!\!\! & \approx & \!\!\!\!\!\!\!\!
   \frac{\sqrt{\pi}}{4\MYhbar^{4}\Gamma_{D_k}} \frac{g_k}{g_i}
   \left[ \frac{d^4_{ik} \mathcal{E}^4_1}{\Delta^4} \right] \Gamma^2_k,
\end{eqnarray}
where $d_{ik}$, $\mathcal{E}_1$, $\Delta$, $\Gamma_k$, $\frac{g_k}{g_i}$, and states $i$ and $k$ are all as defined previously;
the Doppler-broadened state~$k$ linewidth $\frac{\Gamma_{D_k}}{2\pi} \equiv \frac{\nu_{ki}}{c}\sqrt{\frac{k_B T}{M_{\rm Na}}} \approx 1$~GHz;
and again the approximately-equal signs would become exact at their respective orders in perturbation theory in the two equations above
if one assumes that the velocity
distribution of ground-state \ion{Na}{i} atoms within the sodium layer is perfectly Maxwellian, and discounts effects from non-resonant three-photon
processes (which are both very good approximations for the ranges of parameters we consider in this paper)~\citep{Bud08}.

A third additional potentially-significant laser-induced background rate, consisting of the Lorentzian tail of the distribution of
off-resonance transitions to the $3\,$P$_{1/2}$ state, is:
\begin{equation}
\label{Eq:ExRateBkgd3}
W^{\rm Na\,{\scriptscriptstyle I}\,(other\;bkgd.\;excitation\;\#3)}_{(3\,{\rm S}_{1/2}) + \gamma_{\rm 589\,nm} \; \to \; (3\,{\rm P}_{1/2})} \quad \approx \:
  \frac{1}{4\MYhbar^{2}} \frac{g_j}{g_i}
  \left[ \frac{d^2_{ij} \mathcal{E}^2_1}{(\Delta_{3\,{\rm P}_{1/2}})^2} \right] \Gamma_j,
\end{equation}
(again, per ground-state \ion{Na}{i} atom per unit time) where state $j \equiv 3\,$P$_{1/2}$;
the level degeneracy ratio $\frac{g_j}{g_i} = \frac{2}{2} = 1$; $\Delta_{3\,{\rm P}_{1/2}} \equiv (3.2414 \times 10^{12}\:{\rm s}^{-1}) - \Delta$
is the detuning from the $3\,$P$_{1/2}$ state;
the dipole moment $d_{ij} = \sqrt{\frac{3\epsilon_0 h c^3 A_{ji}}{2\omega^3_{ji\phantom{f}}}} = 2.10 \times 10^{-29}$~coulomb-metres;
and both the Einstein $A$ coefficient $A_{ji}$, and the natural linewidth $\Gamma_j$ of state~$j$, are equal to $6.14 \times 10^7$~s$^{-1}$~\citep{Bud08}.

Similar to signal photons, the 589.16~nm photons produced from each of these three background processes will be approximately equal mixtures of
$\hat{x}$ and $\hat{y}$ linear polarizations, and thus the laser and filter polarizations will not have a significant effect in the cases of 
these background sources.  Nor, unfortunately, will the STIRAP laser pulse shapings and timings affect these three backgrounds.
However, as we will calculate in Section~\ref{sec:EstBkgd}, the total rates for each of these three backgrounds are fortunately small in
comparison with the rate for signal.

As we will find in Section~\ref{sec:EstFlux}, our parameters for the 589.16~nm and 819.71~nm lasers
(which we will provide in more detail in Section~\ref{sec:Lasers}) would result in an LPRS
of 11.9 apparent magnitude in both the $r$ and $z$ filters.
And, as we will then calculate in Sections~\ref{sec:ResPrec} and~\ref{sec:EstImp}, such an LPRS would have a major impact on the precision of
dark energy measurements from SNeIa at the Vera C.~Rubin Observatory and at future wide-field SNeIa surveys at other observatories.

\section{Other LPRS Techniques with Detuned Lasers Considered}

We have also considered other atomic and molecular excitations that could potentially form upper-atmospheric light sources with
precise photometric ratios, when utilizing alternative pairs (or triplets) of detuned ground-based lasers.  The constraints on
properties of atomic systems shown in Fig.~\href{https://arxiv.org/pdf/2001.10958.pdf#fig2}{2} in \citetalias{Alb20a}
also apply to the properties of atomic systems with detuned lasers that we consider in this paper, with the important exception of
constraint~1) that is shown at the bottom of that figure 
(under the ``2-laser option'' in that figure).  Thus, we have modified the code
we used in \citetalias{Alb20a}, {\tt LPRSAtomicCascadeFinder},\footnote{Available from the authors upon request.}
to again search the~\citet{NISTtables} database, this time for sets of atomic transitions that obey the required
constraints on atomic systems with detuned ground-based lasers.  In addition to neutral sodium (\ion{Na}{i}), we ran this modified
{\tt LPRSAtomicCascadeFinder} on the same set of tables of upper-atmospheric atomic species from~\citet{NISTtables} that we
considered in \citetalias{Alb20a}
(\ion{Al}{i}, \ion{C}{i}, \ion{Ca}{i}, \ion{Fe}{i}, \ion{H}{i}, \ion{He}{i}, \ion{K}{i}, \ion{N}{i}, \ion{Ne}{i}, \ion{O}{i}, \ion{Al}{ii},
\ion{C}{ii},  \ion{Ca}{ii}, \ion{Fe}{ii}, \ion{H}{ii}, \ion{He}{ii}, \ion{K}{ii}, \ion{N}{ii}, \ion{Na}{ii}, \ion{Ne}{ii}, and \ion{O}{ii}).
The only pair or triplet of atomic transitions that satisfies these required constraints, as coded within the modified {\tt LPRSAtomicCascadeFinder},
is the 589.16~nm and 819.71~nm transitions of \ion{Na}{i} (as shown in Fig.~\ref{fig:LPRSPrimeDiagram} and described in the previous section).
We thus believe that Fig.~\ref{fig:LPRSPrimeDiagram} shows the sole upper-atmospheric {\it atomic} excitation option using a pair (or triplet) of
detuned ground-based lasers that meets the required constraints for such systems.  In addition, we believe that there are no upper-atmospheric
{\it molecular} excitation options (within the optical spectrum)
using pairs or triplets of detuned ground-based lasers, that could provide viable alternative detuned LPRS systems
to the \ion{Na}{i} 589.16~nm and 819.71~nm excitations, either.

\vspace*{-3mm}
\section{Lasers and Launch Telescope}
\label{sec:Lasers}

In this Section we provide an example set of specifications and a design outline for two lasers,
respectively tuned to approximately 3.9~GHz below and above
the \ion{Na}{i} resonances at wavelengths of 589.16~nm and 819.71~nm,
together with a single launch telescope, that would
meet the requirements for an LPRS for precision photometric calibration for the case of the Rubin Observatory.
Due to the order of magnitude lower average laser output powers required (and the resulting far less stringent demands on
the temperature control of internal laser components), these requirements would fortunately be simpler and less costly to engineer
than the case of the single 500~W, 342.78~nm laser LPRS that was considered in \citetalias{Alb20a}.

The maximum optimal laser linewidths $\sigma_\nu$ are determined by the Doppler broadening
$\frac{\nu_0}{c}\sqrt{\frac{k_B T}{M_{\rm Na}}}$ of the $3\,$D$_{5/2}$ excitation ($\nu_0 = 8.75 \times 10^{14}$~Hz)
in the upper atmosphere, where
$T \approx 200$~K.  Thus, the sum of the two laser frequencies should have $\sigma_{\nu} \lesssim 1.5$~GHz;
i.e.~if the two lasers have similar linewidths, then they should each optimally have a maximum linewidth of
$\sigma_{\nu} \; \lesssim \; (1.5$~GHz$)/\sqrt{2} \;\, \approx \; 1$~GHz.

Dye lasers would provide the best performance (i.e., would provide the brightest LPRS with
the highest signal-to-background ratio) if one chooses among unmodified and presently commercially-available laser source options. 
However, as solid-state or fiber laser systems generally tend to be more efficient and have lower maintenance requirements than
dye laser systems, we outline two possible sets of design options below: (A)~Dye laser LPRS design options using a pair of
pulsed dye lasers that are respectively at wavelengths near 589.16~nm and 819.71~nm; and (B)~Solid-state/fiber laser LPRS design options
using the output of an injection-seeded, Q-switched Nd:YAG laser into a pair of optical parametric oscillator (OPO) crystals, for
the generation of pulses at wavelengths near 589.16~nm and 819.71~nm.  As we will show, choices from either one of these sets of
design options would be able to meet the requirements for the generation of an LPRS for high-precision photometric calibration.

We first consider the design options~(A), using a pair of pulsed dye lasers.  A pair of dye lasers such as, for example, either a single Sirah
Double Dye~\citep{Sirah}, or a pair of Radiant Dyes NarrowScan High Repetition Rate lasers~\citep{RadiantDyes}, can produce pairs of pulses
at variable wavelengths respectively near 589.16~nm and 819.71~nm at a repetition rate of 10~kHz, with Gaussian pulses
that are each approximately 5~ns FWHM in length (with each 589~nm pulse trailing each
820~nm pulse by approximately 5~ns $\times \frac{1}{2\sqrt{\ln 2}} \approx 3$~ns for implemention of STIRAP).
In the case of either the Sirah or the Radiant Dyes lasers, each
589~nm pulse can have approximately 1.5~mJ of energy and each 820~nm pulse can have approximately 0.5~mJ of energy (with the
589~nm pulses being approximately a factor of 3 more energetic than the 820~nm pulses because of the approximately $3\times$
greater efficiency of dyes at 589~nm compared with dyes at 820~nm).  The spectral linewidths of
the pulses would be $\le (0.05$~cm$^{-1} = 1.5$~GHz) in the case of both the 589~nm and the 820~nm output light, when double gratings of
approximately (1800~--~2400) lines per mm are used in each of the dye lasers.  In the case of the
Sirah Double Dye, the pair of dye lasers would both be pumped by a single Sirah High Repetition Rate Pulsed Amplifier~\citep{Sirah}
operating at 10 kHz, with its output directed through a high-power polarizing 50:50 beamsplitter cube [such as a
CCM1-PBS25-532-HP/M from~\citet{Thorlabs}], mounted at 45$^{\circ}$ from the polarization axis of the input laser light.  The resulting
pairs of linearly-polarized pulses (each with $>\,2000:1$ extinction ratio between accepted and rejected polarization orientations)
would then respectively pump the 820~nm and 589~nm lasers in the Double Dye laser, however the pump pulses for the 589~nm laser would be
time-delayed with respect to the pump pulses from the 820~nm laser by approximately 3~ns by a variable, and approximately
(60~--~120)~cm, longer light path, prior to entering the Double Dye lasers, to implement STIRAP.  In the case of the pair of Radiant Dyes
lasers, the pair of dye lasers would both be pumped by a single EdgeWave IS-series pulsed green
laser~\citep{EdgeWave} operating at 10~kHz, also with output to a similar polarizing beamsplitter cube, and with the resulting
pair of pump pulses also separately delayed so that the pump pulses entering the 589~nm dye laser arrive approximately 3~ns after
the pump pulses entering the 820~nm dye laser.  Following the pair of either the Radiant Dyes or Sirah dye lasers, the 589~nm and
820~nm output beams would be recombined (and co-aligned) via a dichroic optic [for example product DMLP650L from~\citet{Thorlabs}].
This recombined beam would then be directed to the launch telescope.

Design options~(B), which avoid the use of liquid dyes, could be implemented with, for example, a single 532~nm injection-seeded Amplitude Powerlite
DLS 9050 frequency-doubled Nd:YAG laser~\citep{Amplitude} producing 600~mJ Gaussian pulses that are each approximately 6~ns FWHM in length, with a
spectral linewidth of approximately 0.003~cm$^{-1} = 90$~MHz, at a repetition rate of 50~Hz.  Similar to options~(A) above, the light from this laser   
would be directed through a polarizing 50:50 beamsplitter cube mounted at 45\degr~from the polarization axis of the input
laser light.  The two resulting linearly-polarized beams would each enter separate lithium triborate (LBO) optical parametric oscillator (OPO) crystals. 
The first LBO crystal would be oriented for production of 589.16~nm light (as well as unused 5.48~$\mu$m light), and the second LBO crystal would be
oriented for production of 819.71~nm light (as well as unused 1.52~$\mu$m light).  The wavelengths of the two output beams would be variable between,
respectively, approximately (589.0~--~589.2)~nm and (819.5~--~819.8)~nm, via small adjustments of the angles of the LBO crystals.
The 589~nm pulses would then be time-delayed by approximately 3.6~ns with respect to the 820~nm pulses via an approximately (80~--~140)~cm
longer light path of variable length [similar to the time delay in options~(A) above], to implement STIRAP.  The beams
would then be recombined and co-aligned via a dichroic optic as in options~(A) above, and the
recombined beam would then be directed to the launch telescope.

The launch telescope would maintain the polarization of the two wavelengths, and
expand the combined beam, correspondingly lowering its angular divergence, in order to minimise the resulting beam
diameter at 100~km altitude.  The launch telescope would have the same general optical design as typical launch telescopes for laser guide stars (LGS),
i.e.~expansion of the beam to approximately 0.5~m diameter with the minimum achievable wavefront error.  Also similar to launch telescopes for
LGS (and to the launch telescope considered for the 342.78~nm single-laser LPRS in \citetalias{Alb20a}): as the laser input to the
launch telescope can achieve a beam quality that is within a factor of 2 of diffraction limitation, the resulting output beam from the launch telescope can
achieve an angular divergence that is below 0.2\arcsec~(the pixel scale of the LSST camera at the Rubin Observatory).

As in the single-laser LPRS considered in \citetalias{Alb20a}, the beam diameter at the 100~km altitude of the sodium layer   
will approximately equal the sum in quadrature of the beam diameter at launch telescope exit (0.5~m), the expansion of the beam in the atmosphere
due to its angular divergence at launch telescope exit ($\sim$0.1~m), and the expansion of the beam in the atmosphere due to angular divergence
caused by atmospheric turbulence ($\sim$0.5~m); i.e.~$\sqrt{(0.5)^2 + (0.1)^2 + (0.5)^2}$~m $\approx 0.7$~m, or about 1.4\arcsec~on the sky.
And furthermore just as in the single-laser LPRS considered in \citetalias{Alb20a} (as well as also in LGS), a small additional  
enlargement of the LPRS beam diameter in a radial direction outward from the centre of the telescopic field of view would occur because the centre of the
laser launch telescope would be slightly offset from the centre of the aperture of the observing telescope.  The LPRS will thus be approximately elliptical
in shape on the field of view, with eccentricity of $\sim$0.75 (i.e., the major axis diameter of the LPRS ellipse will be approximately
2.1\arcsec~on the sky, with minor axis diameter being the $\sim$1.4\arcsec~stated above).  Also as in the single-laser LPRS, uncertainties related
to flat-fielding of photometric calibration information across the focal plane of the main telescope could be ameliorated by mounting the launch telescope
to the outer support structure of the main telescope on a tip-tilt stage, so that the launch telescope could tilt up to $\sim1^{\circ}$ in altitude and
azimuth with respect to the main telescope, allowing the LPRS to be moved around the focal plane as needed.

Also as in \citetalias{Alb20a}, we make the assumption that the laser beam spatial profile will be Gaussian, and additionally that the
LPRS profile on the sky will be a Gaussian ellipse.  Although the true LPRS profile on the sky will likely have larger tails
than a Gaussian distribution, the resulting corrections to the analysis in Sections~\ref{sec:EstFlux}~--~\ref{sec:EstImp} 
of this paper from a more detailed (and necessarily more complex) parametrization of the LPRS spatial profile would likely
be fairly small.

\section{Estimation of Observed LPRS Signal Flux}
\label{sec:EstFlux}

We calculate in this Section the expected observed flux at an observatory
that is located at the same
mountaintop site as the pair of source lasers, and launch telescope, each with
properties described in the previous Section, from the resulting 589/590~nm and 819/820~nm
light that is generated by the de-excitation of the $3\,$D$_{5/2}$ (and $3\,$D$_{3/2}$) states of \ion{Na}{i} atoms
in the sodium layer.

{
To use equations~(\ref{Eq:STIRAPGlobalAdiabaticCond})~--~(\ref{Eq:STIRAPExcitationFraction})
to find the total signal excitation rate, we must determine 
the electric field strengths of the two source laser beams $\mathcal{E}_1$ and $\mathcal{E}_2$ within the
altitude range of the sodium layer.
The rms electric field strength, in V$\!/$m, of an electromagnetic plane wave $\mathcal{E} = \sqrt{240 \pi I}$,
where $I$ is the instantaneous intensity in W$\!/$m$^2$.
We have assumed that the laser pulses have Gaussian
spatial and temporal profiles, so if we were to consider the origin in $(x,y,t)$ to be at the
\unskip\parfillskip 0pt \par }

\onecolumn

\noindent centre of a pulse, then the intensity of that pulse (in W$\!/$m$^2$)
$I(x,y,t) = \frac{E}{(2\pi)^{3/2} \sigma_x \sigma_y \sigma_t}
            e^{-\frac{1}{2}\left[ \left( \frac{x}{\sigma_x} \right)^2 + \left( \frac{y}{\sigma_y} \right)^2 +
                                  \left( \frac{t}{\sigma_t} \right)^2 \right]}$,
where $E$ is the energy of a pulse in joules, $t$ and $\sigma_t$ are in units of seconds, and $(x,\sigma_x,y,\sigma_y)$ are all in units of metres.
Per STIRAP, the pulses of the 819.71~nm laser and the 589.16~nm laser will be temporally separated by a time interval~$\tau$.  Defining $\delta_t \equiv 
\frac{\tau}{2}$,   
and centering the origin between pulses from the two lasers, we have that
\begin{eqnarray}
\qquad\mathcal{E}^2_1 \; = \; 240 \pi I_1 \; & = & \; \frac{120 E_1}{(2\pi)^{\frac{1}{2}} \sigma_x \sigma_y \sigma_t}
            e^{-\frac{1}{2}\!\left[ \left( \!\frac{x}{\sigma_x}\! \right)^{\!2} + \left( \!\frac{y}{\sigma_y}\! \right)^{\!2} +
                                    \left( \!\frac{t + \delta_t}{\sigma_t}\! \right)^{\!2} \right]},\;{\rm and}\qquad\qquad\qquad\qquad\qquad\qquad\,\,\\
\qquad\mathcal{E}^2_2 \; = \; 240 \pi I_2 \; & = & \; \frac{120 E_2}{(2\pi)^{\frac{1}{2}} \sigma_x \sigma_y \sigma_t}
            e^{-\frac{1}{2}\!\left[ \left( \!\frac{x}{\sigma_x}\! \right)^{\!2} + \left( \!\frac{y}{\sigma_y}\! \right)^{\!2} +
                                    \left( \!\frac{t - \delta_t}{\sigma_t}\! \right)^{\!2} \right]},\qquad\qquad\qquad\qquad\qquad\qquad\,\,
\end{eqnarray}
and thus
\begin{eqnarray}
\label{Eq:IntE1Field}
\int\limits_{x,y,t}\!\!\!\mathcal{E}^2_1 \; dx \, dy \, dt \; & = & \; 240 \pi E_1 \, ,\;\quad\\
\sqrt{d^2_{ik}\mathcal{E}^2_1 + d^2_{kf}\mathcal{E}^2_2} \; & = & \;
            \sqrt{\frac{120}{(2\pi)^{\frac{1}{2}} \sigma_x \sigma_y \sigma_t}}
            e^{-\frac{1}{4}\!\left[ \! \left( \!\frac{x}{\sigma_x}\! \right)^{\!2} \! + \left( \!\frac{y}{\sigma_y}\! \right)^{\!2} \right]}
            \sqrt{d^2_{ik}E_1 e^{-\frac{1}{2}\!\left( \!\frac{t + \delta_t}{\sigma_t}\! \right)^{\!2}} \!\! +
                  d^2_{kf}E_2 e^{-\frac{1}{2}\!\left( \!\frac{t - \delta_t}{\sigma_t}\! \right)^{\!2}}}\,,\;\quad\\
\label{Eq:IntEFieldProduct}
\qquad\;\frac{1}{\MYhbar}\!\!
\int\limits_{-\infty}^{\infty} \!\!\!\!\left( \! \sqrt{d^2_{ik}\mathcal{E}^2_1 + d^2_{kf}\mathcal{E}^2_2}\, \right) \! dt \; & \approx & \;
                                      \left( \frac{\pi}{2} \right)^{\!\!\frac{1}{4}}\!\!\!
                                      \sqrt{\frac{120\sigma_t}{\sigma_x \sigma_y}}
                                      e^{-\frac{1}{4}\!\left[ \! \left( \!\frac{x}{\sigma_x}\! \right)^{\!2} \! + \left( \!\frac{y}{\sigma_y}\! \right)^{\!2} \right]}\!\!\!
                                      \left[ \!\bigg(\!\! 1 + \Phi \!\!\bigg)\! \!\bigg(\! \eta_1 + \eta_2 \!\bigg)\! +
                                             \!\bigg(\!\! 1 - \Phi \!\!\bigg)\! \!\bigg(\! \frac{\eta^2_1}{2\eta_2} + \frac{\eta^2_2}{2\eta_1} \!\bigg)\!
                                      \right]\!\!,\:\,{\rm and}\quad\quad\\
\label{Eq:IntE1FieldPower4}
\int\limits_{x,y,t} \!\!\! \mathcal{E}^4_1 \; dx \, dy \, dt \; & = & \;
                                      \frac{7200 \sqrt{\pi} E^2_1}{\sigma_x \sigma_y \sigma_t},\;\quad
\end{eqnarray}
where $E_1$ and $E_2$ are, respectively, the 589~nm and the 820~nm
laser pulse energies within the mesosphere in joules; $\eta_1 \equiv \frac{d_{ik}\sqrt{E_1}}{\MYsmhbar}$ and $\eta_2 \equiv \frac{d_{kf}\sqrt{E_2}}{\MYsmhbar}$;
$\Phi \equiv {\rm erf}\!\left( \!\! \frac{\delta_t}{2\sigma_t} \!\! \right) \!$ where
erf($\alpha) \equiv \frac{1}{\sqrt{\pi}}\;\smash{\int\limits^{\alpha}_{}\raisebox{-2.3mm}{$\scriptstyle \!\!\!\!\!\!\!\!\!\!\!\!-\alpha$}} \!e^{-\beta^2}d\beta$
is the typical error function;
and the approximate equality in equation~(\ref{Eq:IntEFieldProduct}) is due to the use of the approximation\footnote{An alternative
approximation to this integral, such as
\begin{align}
\qquad\quad\;\;\;\int\limits_{-\infty}^{\infty} \!\!\! \sqrt{\!\eta^2_1 e^{-\frac{1}{2}\!\left( \!\frac{t + \delta_t}{\sigma_t}\! \right)^{\!2}} \!\! +
                                                  \eta^2_2 e^{-\frac{1}{2}\!\left( \!\frac{t - \delta_t}{\sigma_t}\! \right)^{\!2}}} dt \quad \approx \quad &
\left[ \eta_1 \!\!\!\! \int\limits_{-\infty\phantom{\delta_t}\!}^{-\delta_t}
                                                \!\!\!\! e^{-\frac{1}{4}\! \left( \!\frac{t + \delta_t}{\sigma_t}\! \right)^{\!2}}\!\!\!
                                                     \left( \!\! 1 + \frac{\eta^2_2}{2\eta^2_1}e^{\frac{\delta_t t}{\sigma_t^2}\!} \right)\! dt \right] \! + \!
\left[ \eta_2 \!       \int\limits_{\delta_t}^{\infty}
                                                  \!\!\! e^{-\frac{1}{4}\! \left( \!\frac{t - \delta_t}{\sigma_t}\! \right)^{\!2}}\!\!\!
                                                     \left( \!\! 1 + \frac{\eta^2_1}{2\eta^2_2}e^{-\frac{\delta_t t}{\sigma_t^2}\!} \right)\! dt \right]\nonumber\\
 + & \left[ e^{-\frac{\delta^2_t}{4\sigma^2_t}}\! \sqrt{\eta^2_1 + \eta^2_2} \!  \int\limits_{-\delta_t}^{\delta_t}
                                                  \!\!\! e^{-\frac{t^2}{4\sigma^2_t}}\!\!
                                                     \left( \!\! 1 + \frac{\delta^2_t}{4\sigma^4_t}t^2\! \right)\! dt \right] \! ,\tag{16$^\prime$}
\end{align}
or a numerical solution, could be used instead, but each of those result in a $<\pm10\%$ change in total predicted photon flux
from the approximation in equation~(\ref{Eq:IntApprox}) for all results that we consider.}
\begin{equation}
\label{Eq:IntApprox}
\qquad\int\limits_{-\infty}^{\infty} \!\!\! \sqrt{\!\eta^2_1 e^{-\frac{1}{2}\!\left( \!\frac{t + \delta_t}{\sigma_t}\! \right)^{\!2}} \!\! +
                                                  \eta^2_2 e^{-\frac{1}{2}\!\left( \!\frac{t - \delta_t}{\sigma_t}\! \right)^{\!2}}} dt \quad \approx \quad
\left[ \! \eta_1 \!\!\! \int\limits_{-\infty}^{0} \!\!\!\! e^{-\frac{1}{4}\! \left( \!\frac{t + \delta_t}{\sigma_t}\! \right)^{\!2}}\!\!\!
                                                     \left( \!\! 1 + \frac{\eta^2_2}{2\eta^2_1}e^{\frac{\delta_t t}{\sigma_t^2}\!} \right)\! dt \right] \, + \,
\left[ \! \eta_2 \!\! \int\limits_{0}^{\infty}  \!\!\! e^{-\frac{1}{4}\! \left( \!\frac{t - \delta_t}{\sigma_t}\! \right)^{\!2}}\!\!\!
                                                     \left( \!\! 1 + \frac{\eta^2_1}{2\eta^2_2}e^{-\frac{\delta_t t}{\sigma_t^2}\!} \right)\! dt \right]\!\! .
\end{equation}
First, considering the laser design options~(A) as described in the previous Section, we have that the optical energies within each pulse from the source of the
589~nm and the 820~nm lasers are, respectively, $E_1^{\rm source} = 1.5$~mJ and $E_2^{\rm source} = 0.5$~mJ.  Similar to typical laser guide star systems,
we will assume that $\sim$70\% of that light at each of the two wavelengths is transmitted through the beam transport optics and launch telescope, and projected onto
the sky.  The atmospheric transmission from the 2663~m Cerro Pachon site up to the mesosphere at wavelengths of $\sim$589~nm and $\sim$820~nm is approximately
90\% in the case of both of those wavelengths (with losses dominated by Rayleigh scattering and by water vapor absorption respectively), and thus
$E_1 = (0.9 \times 0.7 \times E_1^{\rm source}) = 945$~$\mu$J and $E_2 = (0.9 \times 0.7 \times E_2^{\rm source}) = 315$~$\mu$J of energy in each pulse
respectively from the two lasers will arrive at the sodium layer.  Since the major and minor axis diameters of the LPRS ellipse are approximately 2.1\arcsec~and
1.4\arcsec~respectively, at the altitude of the sodium layer these diameters will respectively correspond to
approximately 1.1~m and 0.7~m; and considering the diameters to correspond to $\pm1\sigma$ of their respective 1-D Gaussian
distributions, we have that $\sigma_x \approx \frac{1.1\:{\rm m}}{2} = 0.55$~m and
$\sigma_y \approx \frac{0.7\:{\rm m}}{2} = 0.35$~m.  The 5~ns FWHM pulse duration of the
lasers corresponds to Gaussian temporal distributions having
$\sigma_t \approx 2.1$~ns.  Per STIRAP, as discussed in subsection~\ref{subsec:BkgsNeedForSTIRAP},
the temporal separation $\tau = 2\delta_t$ between the pulses will equal $\sqrt{2}\sigma_t$, and thus
$\delta_t \approx 1.5$~ns.
Thus, per equation~(\ref{Eq:IntEFieldProduct}), we have that the pulse area
$\mathcal{A}(x,y) \equiv \displaystyle\frac{\raisebox{-1.5pt}{1}}{\MYhbar}\displaystyle\int_{-\infty}^{\infty} \!\left( \! \sqrt{d^2_{ik}\mathcal{E}^2_1 + d^2_{kf}\mathcal{E}^2_2}\, \right) \! dt
\; \approx \; 24.9 e^{-\frac{1}{4}\!\left[ \! \left( \!\frac{x}{0.55\,{\rm m}}\! \right)^{\!2} \! + \left( \!\frac{y}{0.35\,{\rm m}}\! \right)^{\!2} \right]}$.

Note that we can now define a ``STIRAP pulse area safety factor'' $s_{\!\mathcal{A}} \equiv \frac{\mathcal{A}_{\rm max}}{\mathcal{A}_{\rm min}}$,
where $\mathcal{A}_{\rm min} \equiv 10$ and $\mathcal{A}_{\rm max}$ is the maximum value of $\mathcal{A}(x,y)$, and thus $s_{\!\mathcal{A}} \approx 2.49$ here.
This ``safety factor'' would not directly relate to the brightness of the LPRS, but rather would characterise the degree of concern one might face from
the variety of possible imperfections one might face in the construction and setup of the LPRS system: if $s_{\!\mathcal{A}}$ in the as-built system were to
dip below 1, then STIRAP would not take place in the mesosphere, and thus there would be no LPRS generated

\twocolumn

\noindent at all.  While an expected
value of $s_{\!\mathcal{A}} \approx 2.49$ for an LPRS might seem reasonably safe, higher values of $s_{\!\mathcal{A}}$ would, of course, always be preferable, if possible.

Now we must use equations~(\ref{Eq:STIRAP2GamResonanceWidth}) and~(\ref{Eq:STIRAPExcitationFraction}) to determine the
excitation fraction $f_{\rm STIRAP}$ of \ion{Na}{i} atoms within this mesospheric column that are in the correct velocity range to be excited by the STIRAP process.
The width of the two-photon resonance $\delta_{\nicefrac{1}{2}} = A (\Omega_0)^n (\sigma_t)^{n-1}$, where
$A \approx n \approx 0.9$, $\sigma_t \approx 2.1$~ns,
and $\Omega_0 \approx \frac{d_{ik}}{2\MYsmhbar} \sqrt{\frac{120 E_1}{\sigma_x \sigma_y \sigma_t \sqrt{2\pi}}} = 1.06 \times 10^9$~s$^{-1}$,
and thus $\delta_{\nicefrac{1}{2}} \approx 8.8 \times 10^8$~s$^{-1}$.  As determined within~\ref{subsubsec:DopDet}, the root mean square detuning
$\delta^{\rm RMS}_{\rm eff} \approx 1.02 \times 10^{10}$~s$^{-1}$, and thus $f_{\rm STIRAP} \approx 0.029$.
                                                
Multiplying the elliptical cross-section of the mesospheric column by
this value of $f_{\rm STIRAP}$, by the 10~kHz rate of pulse pairs
from the lasers, and by the column density of approximately $4 \times 10^{13}$ ground-state \ion{Na}{i} atoms per m$^2$, we have
that the total signal excitation (and, thus, total signal de-excitation) rate is
\begin{equation}
\qquad\qquad\;\;\:  2.55 \times 10^{16}\;\ion{Na}{i}\;{\rm atoms \; excited \; per \; second}
\end{equation}
in the mesosphere.

Each of those excited \ion{Na}{i} atoms will emit one 819/820~nm photon, as well as one 589/590~nm photon,
with the photons each emitted in uniform angular distributions.
Again, the atmospheric transmission down to the Cerro Pachon site 95~km below is approximately 90\% in the
case of both of those wavelengths, and thus at the telescope this will correspond to approximately
$N^{\rm{signal}}_{\gamma} \equiv 0.9 \times (2.55 \times 10^{16} \frac{\rm photons}{\rm s}) \times \frac{1}{4\pi \times (9.5 \times 10^{4}\,{\rm m})^{2}} =$
\begin{equation}
\label{Eq:SignalFlux}
\qquad\qquad\qquad\quad\;\;\;\:          2.02 \times 10^5\;{\rm photons/s/m}^2
\end{equation}
at each of 818.55 or 819.70~nm, and 589.16 or 589.76~nm.

Thus, the resulting intensity at the telescope at 818.55/819.70~nm will be approximately $4.9 \times 10^{-14}$~W~m$^{-2}$ $=$ $4.9 \times 10^{-11}$~erg~s$^{-1}$~cm$^{-2}$;
and at 589.16/589.76~nm will be approximately $6.8 \times 10^{-14}$~W~m$^{-2}$ $=$ $6.8 \times 10^{-11}$~erg~s$^{-1}$~cm$^{-2}$.

Following the above, and then using an analogous apparent magnitude calculation as in Section~\href{https://arxiv.org/pdf/2001.10958.pdf#sec6}{6} of
\citetalias{Alb20a}, we have that:
\begin{eqnarray}
m^{r\:{\rm band}}_{\rm AB} & = & 11.9,\\[1mm]
m^{i\:{\rm band}}_{\rm AB} & = & 12.6,\;{\rm and}\\[1mm]
m^{z\:{\rm band}}_{\rm AB} & = & 11.9
\end{eqnarray}
for this LPRS.

If, instead, we consider analogous calculations to the above when using laser design options~(B) rather than design options~(A),
we obtain an expected signal flux of
\begin{equation}
\label{Eq:SignalFluxB}
\qquad\qquad\qquad\quad\;\;\;\:          5.29 \times 10^4\;{\rm photons/s/m}^2
\end{equation}
at each of 818.55 or 819.70~nm, and 589.16 or 589.76~nm; and resulting
intensity at the telescope at 818.55/819.70~nm of approximately $1.3 \times 10^{-14}$~W~m$^{-2}$ $=$ $1.3 \times 10^{-11}$~erg~s$^{-1}$~cm$^{-2}$,
and at 589.16/589.76~nm of approximately $1.8 \times 10^{-14}$~W~m$^{-2}$ $=$ $1.8 \times 10^{-11}$~erg~s$^{-1}$~cm$^{-2}$; and
apparent magnitudes of
\begin{eqnarray}
m^{r\:{\rm band}}_{\rm AB} & = & 13.3,\\[1mm]
m^{i\:{\rm band}}_{\rm AB} & = & 14.0,\;{\rm and}\\[1mm]
m^{z\:{\rm band}}_{\rm AB} & = & 13.3
\end{eqnarray}
for the LPRS.  At first sight, that appears clearly worse (i.e., almost a factor of 4 less bright) when compared with laser design options~(A),
however note that
the ``STIRAP pulse area safety factor'' $s_{\!\mathcal{A}}$ that we defined above would be approximately equal to
58.2 for laser design options~(B), i.e.~over 20 times greater than the value of $s_{\!\mathcal{A}} \approx 2.49$ for laser design options~(A);
and also that
both of the two design options appear to be sufficiently bright for use as an LPRS at surveys performed by large telescopes (such as at the Rubin Observatory).
Also note that laser design options~(B) could potentially be made as bright as design options~(A), without sacrificing their additional safety factor, if the
laser pulse repetition rate could be increased significantly beyond 50~Hz.  [Analogously, the safety factor
$s_{\!\mathcal{A}}$ could be increased in laser design options~(A), toward the high value of $s_{\!\mathcal{A}}$ that is expected in
laser design options~(B), if the energies of the individual dye laser output pulses could be increased significantly.]

Table~\ref{tab:ExpSig} summarises the expected numbers of signal photons and resulting observed signal photoelectrons for the case
of the Simonyi Survey Telescope at the Rubin Observatory [when using the expected telescope, camera, and filter throughputs as documented in~\citet{Jones19},
with unmodified telescope $r$, $i$, and $z$ filters], in the cases of either laser design options~(A) or~(B).
The table rows containing the the total numbers of detected LPRS signal photoelectrons, as well as the table rows below them and
the analysis in Sections~\ref{sec:ResPrec} and~\ref{sec:EstImp} of this paper, include the important effect of the loss of 55\% of the signal
photons due to the necessary inclusion of linear polarization filters within the telescope's optical filters to reject Rayleigh-scattered
background light.

Similarly to the single-laser LPRS
described in~\citetalias{Alb20a}, the dominant systematic uncertainty on the predicted 1:1
ratio between the photon flux at 589/590~nm vs.~819/820~nm of this two-laser LPRS would be due to the possibility
of inelastic collisions of excited \ion{Na}{i} atoms in the mesosphere during the very brief period of atomic de-excitation.
However, unlike the single-laser LPRS, this two-laser LPRS has associated background light that will slightly
modify the central value of the 1:1 ratio, instead of the expectation value of the ratio being exactly 1:1 and there only
being an associated systematic uncertainty on that value.  The
predicted ratio of photon flux at 589/590~nm vs.~819/820~nm, including both its central value modification and its
systematic uncertainty, will thus equal $((1 + \delta) \pm \epsilon):1$.
As we will show in the following Section, the predicted central value modification $\delta$ will equal
approximately $2.0 \times 10^{-5}$ for the case of laser design options~(A).  Similarly to the single-laser LPRS,
we conservatively estimate the systematic uncertainty on the ratio $\epsilon$ to be $9 \times 10^{-5}$ for this two-laser LPRS.

\vspace*{-5mm}
\section{Estimation of Observed LPRS Background}
\label{sec:EstBkgd}

The laser-induced background light associated with this LPRS can be divided into the following six categories:
\begin{enumerate}[label=\bfseries{[}{\arabic*}{]}, leftmargin = 5mm, labelsep = *]
\item Background photons from the de-excitations of single-photon transition events of ground-state \ion{Na}{i} to the $3\,$P$_{3/2}$ state 
(i.e., ``other background excitation~\#1''), with rate as described by equation~(\ref{Eq:ExRateBkgd1});
\item Background photons from the de-excitations of three-photon transition events of ground-state \ion{Na}{i} to the $3\,$P$_{3/2}$ state
(i.e., ``other background excitation~\#2''), with rate as described by equation~(\ref{Eq:ExRateBkgd2});
\item Background photons from the de-excitations of two-photon transition events of ground-state \ion{Na}{i} to the $3\,$P$_{1/2}$ state
(i.e., ``other background excitation~\#3''), with rate as described by equation~(\ref{Eq:ExRateBkgd3});
\item Background photons from the de-excitations of \ion{Na}{i} virtual $3\,$P$^{*}_{3/2}$ excitation production, with rate
as described by equation~(\ref{Eq:DeExRateBkgdNoSTIRAP}), that fail to be completely eliminated by both STIRAP and polarization filtering;
\end{enumerate}

\onecolumn
\begin{table*}
\begin{minipage}{10cm}
\caption{\label{tab:ExpSig}Expected LPRS signal photon flux and photoelectrons collected at the telescope.}
\end{minipage}
\begin{center}
\begin{tabular}{lll}
\hline \hline
\rule{0mm}{4mm}                         & \hspace*{10.8mm}\textbf{Expected value}    & \hspace*{5.8mm}\textbf{Expected value}         \\
\raisebox{1.5ex}[0pt]{\hspace*{-1mm}
\textbf{Quantity}}                      & \raisebox{0.3ex}[0pt]{
                                          \hspace*{4mm}for laser design options~(A)} & \raisebox{0.3ex}[0pt]{for laser design options~(B)} \\[-0.3mm]
\hline
\textbf{LPRS signal photon flux} at the telescope, \rule{0mm}{4.5mm} &             &                                                  \\
at each of 589/590~nm and 819/820~nm                      &   \raisebox{1.5ex}[0pt]{\hspace*{8.6mm}$2.02 \times 10^5$~photons/s/m$^2$} & 
                                                              \raisebox{1.5ex}[0pt]{\hspace*{0.9mm}$5.29 \times 10^4$~photons/s/m$^2$}\\
\textbf{Total LPRS signal photon collection rate}  \rule{0mm}{6mm}   &             &                                                  \\
at each of 589/590~nm and 819/820~nm                      &                        &                                                  \\
(within the 35~m$^2$ clear aperture of the                &            \hspace*{7.8mm} $7.07 \times 10^6$~photons/s                    &
                                                                       \hspace*{0.1mm} $1.85 \times 10^6$~photons/s                   \\
Simonyi Survey Telescope at the Rubin                     &                        &                                                  \\
Observatory)                                              &                        &                                                  \\
\textbf{Total number of detected LPRS signal}      \rule{0mm}{6mm}   & $r$ filter:\hspace*{0.5mm}$9.13 \times 10^7$~photoelectrons
                                                                     & \hspace*{0.7mm}$2.39 \times 10^7$~photoelectrons               \\
\textbf{photoelectrons} within the elliptical                        & $i$\hspace*{0.5mm} filter:\hspace*{0.5mm}$3.82 \times 10^7$~photoelectrons
                                                                     & \hspace*{0.7mm}$1.00 \times 10^7$~photoelectrons               \\
LPRS spot during a 30~s visit                                        & $z$ filter:\hspace*{0.7mm}$5.94 \times 10^7$~photoelectrons
                                                                     & \hspace*{0.7mm}$1.56 \times 10^7$~photoelectrons               \\
\textbf{Detected signal photoelectrons per} \rule{0mm}{6mm}          & $r$ filter:\hspace*{0.5mm}$1.52 \times 10^6$~photoelectrons
                                                                     & \hspace*{0.7mm}$3.99 \times 10^5$~photoelectrons               \\
\textbf{\boldmath 0.2$^{\prime\prime} \times$ 0.2$^{\prime\prime}$ pixel} at the centre of 
                                                                     & $i$\hspace*{0.5mm} filter:\hspace*{0.5mm}$6.40 \times 10^5$~photoelectrons
                                                                     & \hspace*{0.7mm}$1.68 \times 10^5$~photoelectrons               \\
the LPRS spot during a 30~s visit                                    & $z$ filter:\hspace*{0.7mm}$9.92 \times 10^5$~photoelectrons
                                                                     & \hspace*{0.7mm}$2.60 \times 10^5$~photoelectrons               \\
\textbf{Signal standard deviation per 0.2$^{\prime\prime} \times$ 0.2$^{\prime\prime}$}  \rule{0mm}{6mm}    
                                                                     & $r$ filter:\hspace*{0.5mm}$1.23 \times 10^3$~photoelectrons
                                                                     & \hspace*{0.7mm}$6.32 \times 10^2$~photoelectrons               \\
\textbf{pixel} at the centre of the LPRS spot during                 & $i$\hspace*{0.5mm} filter:\hspace*{0.5mm}$8.00 \times 10^2$~photoelectrons
                                                                     & \hspace*{0.7mm}$4.10 \times 10^2$~photoelectrons               \\
a 30~s visit                                                         & $z$ filter:\hspace*{0.7mm}$9.96 \times 10^2$~photoelectrons
                                                                     & \hspace*{0.7mm}$5.10 \times 10^2$~photoelectrons               \\
\hline
\end{tabular}
\end{center}
\end{table*}

\begin{enumerate}[label=\bfseries{[}{\arabic*}{]}, leftmargin = 5mm, labelsep = *, resume]
\item Background photons from near-180\degr~atmospheric Rayleigh back-scattering; and
\item Background photons from near-180\degr~atmospheric Raman back-scattering and
de-excitation light from other inelastic excitations.
\end{enumerate}

We calculate the expected amount of background light from each of these categories in the following paragraphs.
As we will show, background light from category~[5] will be the dominant category of laser-induced
background light.\vspace*{-4mm}

\subsection{Laser-induced background category~[1]} 
The expected amount of category~[1] of laser-induced background light can be calculated using
equations~(\ref{Eq:ExRateBkgd1}) and~(\ref{Eq:IntE1Field}).
This excitation cross-section, per 589.16~nm laser pulse and per ground-state \ion{Na}{i} atom, 
${\displaystyle \int_{x,y,t}}\!\!\left(
 W^{\rm Na\,{\scriptscriptstyle I}\,(other\;bkgd.\;excitation\;\#1)}_{(3\,{\rm S}_{1/2}) + \gamma_{\rm 589\,nm} \; \to \; (3\,{\rm P}_{3/2})}  
 \right) \! dx \, dy \, dt \;\; \approx$\\
\hspace*{5.5cm} $(1.03 \times 10^{-6})$~m$^2 \:\:$ in the case of laser design options~(A), and\\*[1mm]
\hspace*{5.5cm} $(3.43 \times 10^{-4})$~m$^2 \:\:$ in the case of laser design options~(B).

Thus, the flux $N^{{\rm cat.\:[1]\:bkgd.}}_{\gamma}$ at the telescope from this background category will be approximately:
\begin{center}
$(1.03 \times 10^{-6})\frac{{\rm m}^2}{\rm (pulse)(Na\:{\scriptscriptstyle I}\:atom)} \times (1 \times 10^4)\frac{\rm pulses}{\rm s} \times (4 \times 10^{13})\frac{\rm Na\:{\scriptscriptstyle I}\:atoms}{{\rm m}^2} 
\times \frac{0.9}{4\pi \times (9.5 \times 10^{4}\:{\rm m})^{2}} \quad = \quad 3.26\,\frac{\rm (589\:nm\:photons)}{\rm s\:m^2}$
\end{center}
for laser design options~(A), and:
\begin{center}
$(3.43 \times 10^{-4})\frac{{\rm m}^2}{\rm (pulse)(Na\:{\scriptscriptstyle I}\:atom)} \times (5 \times 10^1)\frac{\rm pulses}{\rm s} \times (4 \times 10^{13})\frac{\rm Na\:{\scriptscriptstyle I}\:atoms}{{\rm m}^2} 
\times \frac{0.9}{4\pi \times (9.5 \times 10^{4}\:{\rm m})^{2}} \quad = \quad 5.44\,\frac{\rm (589\:nm\:photons)}{\rm s\:m^2}$
\end{center}
for laser design options~(B).\vspace*{-3mm}

\subsection{Laser-induced background category [2]}

We use equations~(\ref{Eq:ExRateBkgd2}) and~(\ref{Eq:IntE1FieldPower4}) to calculate the expected
amount of background light from category~[2].  The cross-section per 589.16~nm laser pulse and per ground-state \ion{Na}{i} atom
${\displaystyle \int_{x,y,t}}\!\!\left(
 W^{\rm Na\,{\scriptscriptstyle I}\,(other\;bkgd.\;excitation\;\#2)}_{(3\,{\rm S}_{1/2}) + 2\gamma_{\rm 589\,nm} \; \to \; (3\,{\rm P}_{3/2}) + \gamma_{\rm 589\,nm}}
 \right) \! dx \, dy \, dt \;\; \approx$\\
\hspace*{5.5cm} $(6.19 \times 10^{-8})$~m$^2 \:\:$ in the case of laser design options~(A), and\\*[1mm]
\hspace*{5.5cm} $(5.68 \times 10^{-4})$~m$^2 \:\:$ in the case of laser design options~(B).

Thus the flux $N^{{\rm cat.\:[2]\:bkgd.}}_{\gamma}$ at the telescope from this background category will be approximately:
\begin{center}
$(6.19 \times 10^{-8})\frac{{\rm m}^2}{\rm (pulse)(Na\:{\scriptscriptstyle I}\:atom)} \times (1 \times 10^4)\frac{\rm pulses}{\rm s} \times (4 \times 10^{13})\frac{\rm Na\:{\scriptscriptstyle I}\:atoms}{{\rm m}^2} 
\times \frac{0.9}{4\pi \times (9.5 \times 10^{4}\:{\rm m})^{2}} \quad = \quad 0.20\,\frac{\rm (589\:nm\:photons)}{\rm s\:m^2}$ 
\end{center}
for laser design options~(A), and:
\begin{center}
$(5.68 \times 10^{-4})\frac{{\rm m}^2}{\rm (pulse)(Na\:{\scriptscriptstyle I}\:atom)} \times (5 \times 10^1)\frac{\rm pulses}{\rm s} \times (4 \times 10^{13})\frac{\rm Na\:{\scriptscriptstyle I}\:atoms}{{\rm m}^2}
\times \frac{0.9}{4\pi \times (9.5 \times 10^{4}\:{\rm m})^{2}} \quad = \quad 9.01\,\frac{\rm (589\:nm\:photons)}{\rm s\:m^2}$
\end{center}
for laser design options~(B).\vspace*{-3mm}

\subsection{Laser-induced background category [3]}

We use equations~(\ref{Eq:ExRateBkgd3}) and~(\ref{Eq:IntE1Field}) to calculate the expected
amount of background light from category~[3].  The cross-section per 589.16~nm laser pulse and per ground-state \ion{Na}{i} atom
${\displaystyle \int_{x,y,t}}\!\!\left(
 W^{\rm Na\,{\scriptscriptstyle I}\,(other\;bkgd.\;excitation\;\#3)}_{(3\,{\rm S}_{1/2}) + \gamma_{\rm 589\,nm} \; \to \; (3\,{\rm P}_{1/2})}
 \right) \! dx \, dy \, dt \;\; \approx$\\
\hspace*{5.5cm} $(4.19 \times 10^{-8})$~m$^2 \:\:$ in the case of laser design options~(A), and\\*[1mm]
\hspace*{5.5cm} $(1.40 \times 10^{-5})$~m$^2 \:\:$ in the case of laser design options~(B).

Thus the flux $N^{{\rm cat.\:[3]\:bkgd.}}_{\gamma}$ at the telescope from this background category will be approximately:
\begin{center}
$(4.19 \times 10^{-8})\frac{{\rm m}^2}{\rm (pulse)(Na\:{\scriptscriptstyle I}\:atom)} \times (1 \times 10^4)\frac{\rm pulses}{\rm s} \times (4 \times 10^{13})\frac{\rm Na\:{\scriptscriptstyle I}\:atoms}{{\rm m}^2}
\times \frac{0.9}{4\pi \times (9.5 \times 10^{4}\:{\rm m})^{2}} \quad = \quad 0.13\,\frac{\rm (589\:nm\:photons)}{\rm s\:m^2}$ 
\end{center}
for laser design options~(A), and:
\begin{center}
$(1.40 \times 10^{-5})\frac{{\rm m}^2}{\rm (pulse)(Na\:{\scriptscriptstyle I}\:atom)} \times (5 \times 10^1)\frac{\rm pulses}{\rm s} \times (4 \times 10^{13})\frac{\rm Na\:{\scriptscriptstyle I}\:atoms}{{\rm m}^2}
\times \frac{0.9}{4\pi \times (9.5 \times 10^{4}\:{\rm m})^{2}} \quad = \quad 0.22\,\frac{\rm (589\:nm\:photons)}{\rm s\:m^2}$
\end{center}
for laser design options~(B).

\subsection{Laser-induced background category [4]}

The expected amount of background light from category~[4] can be calculated using
equations~(\ref{Eq:DeExRateBkgdNoSTIRAP}) and~(\ref{Eq:IntE1Field}).  The cross-section per 589.16~nm laser pulse and per ground-state \ion{Na}{i} atom 
${\displaystyle \int_{x,y,t}}\!\!\left(
 W^{\rm Na\,{\scriptscriptstyle I}\,(bkgd.\;de\mhyphen excitation\;during\;pulse,\;if\;no\;STIRAP)}_{(3\,{\rm P}^{*}_{3/2}) \; \to \; (3\,{\rm S}_{1/2}) + \gamma_{\rm 589\,nm}}
 \right) \! dx \, dy \, dt \;\; \approx$\\
\hspace*{5.5cm} $(1.46 \times 10^{-3})$~m$^2 \:\:$ in the case of laser design options~(A), and\\*[1mm]
\hspace*{5.5cm} $(4.88 \times 10^{-1})$~m$^2 \:\:$ in the case of laser design options~(B).\\*[1mm]
However, this is, of course, the cross-section if STIRAP were not used; whereas we are utilizing STIRAP.
Thus, we must multiply this without-STIRAP cross-section by the expected small fraction of nonadiabatic losses from the STIRAP process.  In
well-controlled laboratory experiments, the nonadiabatic loss fraction from STIRAP has been reduced to levels
within the range $(10^{-6})$ -- $(10^{-8})$ using carefully-shaped pulses and other optimizations~\citep{Vit17}.
With our presently-considered application of STIRAP from mountaintop-located lasers to the
open upper atmosphere of Earth, we conservatively make the assumption of a nonadiabatic loss fraction at the $10^{-4}$ level,
and thus a STIRAP-modified effective cross-section of\\
\hspace*{5.5cm} $(1.46 \times 10^{-7})\frac{{\rm m}^2}{\rm (pulse)(Na\:{\scriptscriptstyle I}\:atom)} \:\:$ in the case of laser design options~(A), and\\*[1mm]
\hspace*{5.5cm} $(4.88 \times 10^{-5})\frac{{\rm m}^2}{\rm (pulse)(Na\:{\scriptscriptstyle I}\:atom)} \:\:$ in the case of laser design options~(B).

Thus, we estimate the flux $N^{{\rm cat.\:[4]\:bkgd.}}_{\gamma}$ at the telescope from this background category to be approximately:
\begin{center}
$(1.46 \times 10^{-7})\frac{{\rm m}^2}{\rm (pulse)(Na\:{\scriptscriptstyle I}\:atom)} \times (1 \times 10^4)\frac{\rm pulses}{\rm s} \times (4 \times 10^{13})\frac{\rm Na\:{\scriptscriptstyle I}\:atoms}{{\rm m}^2} 
\times \frac{0.9}{4\pi \times (9.5 \times 10^{4}\:{\rm m})^{2}} \quad = \quad 0.47\,\frac{\rm (589\:nm\:photons)}{\rm s\:m^2}$
\end{center}
for laser design options~(A), and:
\begin{center}
$(4.88 \times 10^{-5})\frac{{\rm m}^2}{\rm (pulse)(Na\:{\scriptscriptstyle I}\:atom)} \times (5 \times 10^1)\frac{\rm pulses}{\rm s} \times (4 \times 10^{13})\frac{\rm Na\:{\scriptscriptstyle I}\:atoms}{{\rm m}^2}
\times \frac{0.9}{4\pi \times (9.5 \times 10^{4}\:{\rm m})^{2}} \quad = \quad 0.78\,\frac{\rm (589\:nm\:photons)}{\rm s\:m^2}$ 
\end{center}
for laser design options~(B).

\subsection{Laser-induced background category [5]}

The expected amount of Rayleigh-backscattered laser light that both enters the telescope aperture and is
superimposed over the LPRS spot can be calculated using the same technique as in Section~\href{https://arxiv.org/pdf/2001.10958.pdf#sec7}{7} of 
\citetalias{Alb20a} [and specifically using
equation~(\href{https://arxiv.org/pdf/2001.10958.pdf#eq12}{12}) in that Section, 
modified for the different laser wavelengths used in the present paper].  The analogous fractions
\begin{eqnarray}
\qquad\qquad f_{b, {\rm over\:LPRS\:spot}}^{R,\gamma_\lambda}
      & \approx & \frac{\scriptstyle 20\pi \times (3.6 \times 10^{-31}) \times (\lambda_\gamma)^{-4.0117}}{\scriptstyle 3}
                  \int^{105\,000}_{z = 80\,000} e^{-\left( \frac{z}{8800} \right)} dz
                  \int^{\tan^{-1}(\frac{r}{z - 2663})}_{\theta = 0} \sin\theta d\theta                                                                                 \\
      & \approx & 7.96 \times 10^{-14}\;\rm{(for}\:\lambda_\gamma = 589.16\:\rm{nm)\quad and \quad}2.12 \times 10^{-14}\;\rm{(for}\:\lambda_\gamma = 819.71\:\rm{nm)}, \nonumber
\end{eqnarray}
and analogous total numbers of laser photons reaching the mesosphere per second\\*[1mm]
\hspace*{3.1cm} $N^{\rm{589\,nm\,laser}}_{\gamma_{\rm 589\,nm}} = 2.80 \times 10^{19} \:\:$ and 
$\:\: N^{\rm{820\,nm\,laser}}_{\gamma_{\rm 820\,nm}} = 1.30 \times 10^{19} \:\:$ in the case of laser design options~(A); and\\*[1mm]
\hspace*{3.1cm} $N^{\rm{589\,nm\,laser}}_{\gamma_{\rm 589\,nm}} = 4.67 \times 10^{19} \:\:$ and 
$\:\: N^{\rm{820\,nm\,laser}}_{\gamma_{\rm 820\,nm}} = 3.90 \times 10^{19} \:\:$ in the case of laser design options~(B).\\*[1mm]
Thus, the two true fluxes at the telescope from this background category will be:\\*[1mm]
\hspace*{3.1cm} $f_{b, {\rm over\:LPRS\:spot}}^{R,\gamma_{\rm 589\,nm}} N^{\rm{589\,nm\,laser}}_{\gamma_{\rm 589\,nm}} \quad \approx \quad (2.23 \times 10^6)\,\frac{\rm (589\:nm\:photons)}{\rm s\:m^2}$ and\\*[1mm]
\hspace*{3.1cm} $f_{b, {\rm over\:LPRS\:spot}}^{R,\gamma_{\rm 820\,nm}} N^{\rm{820\,nm\,laser}}_{\gamma_{\rm 820\,nm}} \quad \approx \quad (2.75 \times 10^5)\,\frac{\rm (820\:nm\:photons)}{\rm s\:m^2}$\\*[2mm]
in the case of laser design options~(A); and

\twocolumn

\noindent\hspace*{0.1cm} $f_{b, {\rm over\:LPRS\:spot}}^{R,\gamma_{\rm 589\,nm}} N^{\rm{589\,nm\,laser}}_{\gamma_{\rm 589\,nm}} \quad \approx \quad (3.71 \times 10^6)\,\frac{\rm (589\:nm\:photons)}{\rm s\:m^2}$ and\\*[1mm]
\hspace*{0.1cm} $f_{b, {\rm over\:LPRS\:spot}}^{R,\gamma_{\rm 820\,nm}} N^{\rm{820\,nm\,laser}}_{\gamma_{\rm 820\,nm}} \quad \approx \quad (8.26 \times 10^5)\,\frac{\rm (820\:nm\:photons)}{\rm s\:m^2}$\\*[1mm]
in the case of laser design options~(B).\\

However, the 589~nm photons from this background category will be predominantly $\hat{x}$-polarized
and the 820~nm photons from this background category will be predominantly $\hat{y}$-polarized,
and thus they will respectively be
blocked by the telescope $r$ filter (for the 589~nm photons), and by the telescope $i$ and $z$ filters (for the 820~nm photons).
Thus there are three ways that a photon from this
background category could get
through the telescope $r$, $i$, or $z$ filter:
\vspace*{2mm}
\begin{enumerate}[label=({\arabic*}), leftmargin = 5mm, labelsep = *, itemsep = 0mm, topsep = 0mm]
\item Imperfect polarization of the output of the lasers, resulting in an admixture of imperfectly-polarized photons from the source; 
\item Depolarization of the polarized photons within the atmosphere, either during upward or downward transit; or 
\item Imperfect rejection of properly-polarized background photons by the telescope filters.
\end{enumerate}
\vspace*{2mm}
For (1), the dominant cause of imperfect polarization from the laser source would be from imperfectly-polarized output of
the polarizing beamsplitter cube.
Polarizing beamsplitters such as the CCM1-PBS25-532-HP/M~\citep{Thorlabs}
that we considered in Section~\ref{sec:Lasers} advertise a greater than 1000:1 ratio between accepted and rejected polarizations, so we
conservatively estimate at the bottom of this range, i.e.~a $\frac{1}{1000}$ admixture of incorrectly-polarized photons from
the laser sources.  For (2), typical depolarization fractions of a laser beam following a vertical path through the Earth's atmosphere
are in the range of $(1 \times 10^{-7}) - (5 \times 10^{-5})$~\citep{Hohn69}, thus this effect would be relatively negligible.  For (3),
the rejection of properly-polarized background photons would be implemented using a
polarizing filter, which should have a greater than 1000:1 ratio between accepted and rejected polarizations.
Again, we conservatively estimate at the bottom of this range, and thus we estimate that $\frac{1}{1000}$ of incident properly-polarized
photons will manage to pass through the telescope filters.  The combination of (1), (2), and (3) will approximately result in the
sum of the three effects, and thus the true fluxes
from this background category should be
multiplied by a factor of approximately $\frac{1}{1000} + \frac{1}{1000} = \frac{1}{500}$ due to the rejection of properly-polarized
photons from this background category, in order to obtain the resulting effective fluxes.

Thus, the effective fluxes at the telescope from this background category will be approximately:
\begin{center}
$\frac{1}{500} \times f_{b, {\rm over\:LPRS\:spot}}^{R,\gamma_{\rm 589\,nm}} N^{\rm{589\,nm\,laser}}_{\gamma_{\rm 589\,nm}} \;\: \approx \;\: (4.46 \times 10^3)\,\frac{\rm (589\:nm\:photons)}{\rm s\:m^2}$ 
\end{center}
and
\begin{center}
$\frac{1}{500} \times f_{b, {\rm over\:LPRS\:spot}}^{R,\gamma_{\rm 820\,nm}} N^{\rm{820\,nm\,laser}}_{\gamma_{\rm 820\,nm}} \;\: \approx \;\: (5.50 \times 10^2)\,\frac{\rm (820\:nm\:photons)}{\rm s\:m^2}$
\end{center}
in the case of laser design options~(A); and
\begin{center}
$\frac{1}{500} \times f_{b, {\rm over\:LPRS\:spot}}^{R,\gamma_{\rm 589\,nm}} N^{\rm{589\,nm\,laser}}_{\gamma_{\rm 589\,nm}} \;\: \approx \;\: (7.42 \times 10^3)\,\frac{\rm (589\:nm\:photons)}{\rm s\:m^2}$ 
\end{center}
and
\begin{center}
$\frac{1}{500} \times f_{b, {\rm over\:LPRS\:spot}}^{R,\gamma_{\rm 820\,nm}} N^{\rm{820\,nm\,laser}}_{\gamma_{\rm 820\,nm}} \;\: \approx \;\: (1.65 \times 10^3)\,\frac{\rm (820\:nm\:photons)}{\rm s\:m^2}$
\end{center}
in the case of laser design options~(B).

This seems like it would result in a major problem, since these fluxes from this background category are a significant fraction of
the expected signal fluxes from equations~(\ref{Eq:SignalFlux}) and~(\ref{Eq:SignalFluxB}), even after the above small wrong-polarization acceptance factor of
approximately~$\frac{1}{500}$ for this background is included.  However, unlike background from categories~[1]~--~[4], the Rayleigh-scattered background
will not be superimposed purely on the LPRS spot, but rather will form a continuous streak, as was discussed
in Section~\href{https://arxiv.org/pdf/2001.10958.pdf#sec7}{7} of \citetalias{Alb20a}; and thus one is able to perform a combined fit to
the Rayleigh streak as a continuous (and approximately exponentially-falling)
distribution within the telescope camera images, together with a fit to the LPRS spot that rests on top of the tail of that continuous distribution.
This fit technique will be discussed further, and demonstrated, in the following Section on expected photometric ratio precision.
One is, thus, still able to achieve high precision on the observationally-fitted photometric ratio.

\subsection{Laser-induced background category [6]}

Similarly to the case of the 342.78~nm laser in \citetalias{Alb20a},
the atmospheric Raman backscattering of light from the 589.16~nm and 819.71~nm lasers in the present paper would contribute to the
observed LPRS background.  However, just as in \citetalias{Alb20a}, the largest such background contributions would be from the strong Raman lines in the
Schumann-Runge bands of O$_2$, which will produce cross-sections only of order $10^{-40}$~cm$^2$ per molecule, and are thus also negligible here.

As also noted in \citetalias{Alb20a}, the maximum line intensity of O$_2$ and N$_2$ Raman rotational transitions corresponding to the first vibrational
excitation in these molecules is approximately $10^{-16}$~erg~s$^{-1}$~cm$^{-2}$~\citep{Cal14,Vog17}.  That value is negligible when compared with the returned signal 
flux from the sodium layer, which will have an intensity at 589.16/589.76~nm (calculated in Section~\ref{sec:EstFlux} of this paper) of approximately 
$7 \times 10^{-11}$~erg~s$^{-1}$~cm$^{-2}$ in the case of laser design options~(A), and approximately $2 \times 10^{-11}$~erg~s$^{-1}$~cm$^{-2}$ 
in the case of laser design options~(B).

\subsection{Total laser-induced background}

The laser-induced backgrounds from the six different categories will all interfere
with one another, and thus one must consider their relative phases in order to add them.  However, in practice, the contributions from the different
background categories will not be coherent with one another over timescales longer than at most a few tens of nanoseconds, and in any case the background
from category~[5] greatly dominates over
the other categories; thus one can approximate the total laser-induced background either by the incoherent sum of light from the six categories, or by just
the light from category~[5]; those two methods produce essentially the same result.  Thus, we estimate the total effective laser-induced background fluxes
from within the 2.1\arcsec~$\times$~1.4\arcsec~diameter LPRS ellipse to be
$(4.5 \times 10^3)\,\frac{\rm (589\:nm\:photons)}{\rm s\:m^2}$ and
$(5.5 \times 10^2)\,\frac{\rm (820\:nm\:photons)}{\rm s\:m^2}$ in the case of laser design options~(A); and
$(7.4 \times 10^3)\,\frac{\rm (589\:nm\:photons)}{\rm s\:m^2}$ and
$(1.7 \times 10^3)\,\frac{\rm (820\:nm\:photons)}{\rm s\:m^2}$ in the case of laser design options~(B).   

However, as noted above, the dominant background from category~[5] is of a different
nature than the other background categories, in that it will form a continuous streak within each image; and thus one can perform an
image-by-image combined fit for category~[5] background as well as for signal.  Whereas the background from other categories,
with effective fluxes that per our estimations will sum to a total of
$N^{\rm{peaking\:bkgd.}}_{\gamma} \equiv \sum\limits_{i=1}^4 N^{{\rm cat.\:[}i{\rm ]\:bkgd.}}_{\gamma} = 4.04\,\frac{\rm (589\:nm\:photons)}{\rm s\:m^2}$
in the case of laser design options~(A) and $15.45\,\frac{\rm (589\:nm\:photons)}{\rm s\:m^2}$ in the case of laser design options~(B), will be
superimposed on the LPRS spot; and thus one must use other techniques besides image-by-image fitting --- for example, multiple-image scans with
variation of the laser detuning parameter $\Delta$, and/or of the relative intensities, and/or pulse shapes, of the two lasers --- to
experimentally determine those fluxes.  We additionally note that the expected central value modification $\delta$ on the 1:1 ratio between
the photon flux at 589/590~nm vs.~819/820~nm that was mentioned at the end of Section~\ref{sec:EstFlux} will equal
$\left. \left( N^{\rm{peaking\:bkgd.}}_{\gamma} \right) \middle/ \left( N^{\rm{signal}}_{\gamma} \right) \right. = \frac{4.04}{2.02 \times 10^5} = 2.0 \times 10^{-5}$
in the case of laser design options~(A), and $\frac{15.45}{5.29 \times 10^4} = 2.9 \times 10^{-4}$ in the case of laser design options~(B);
where $N^{\rm{signal}}_{\gamma}$ is the signal photon flux from equation~(\ref{Eq:SignalFlux}) in the case of laser design options~(A), and
from equation~(\ref{Eq:SignalFluxB}) in the case of laser design options~(B).

\subsection{Non-laser-induced backgrounds}

In addition to the above total amount
of laser-induced background light calculated in the previous paragraphs,
there will also be the typical diffuse sky background (including all other [i.e., non-laser]
light that is scattered by the optical elements of the telescope, by the atmosphere, and by zodiacal dust),
as well as instrumental background noise.  These sources of diffuse background will be corrected through the usual
technique of sky subtraction.  The amount of diffuse sky background at the Rubin Observatory site is estimated in~\citet{Jones17} to be,
for the $r$, $i$, and $z$ filters respectively:
\begin{itemize}[label = \textbullet, leftmargin = 4mm, labelsep = *, itemsep = 1mm, topsep = 1mm]
\item An apparent magnitude per square arcsecond of 21.2, 20.5, and 19.6;
\item 42.8, 61.5, and 101.3~$\frac{\rm photons}{\rm s\:m^{2}\:(square\:arcsecond)}$;
\item 1498, 2151, and 3544~photons/s/(square arcsecond) within the 35~m$^2$ clear aperture of the Simonyi Survey Telescope at the Rubin Observatory;
\item 3459, 5141, and 8472~photons/s from within the elliptical LPRS spot of angular diameter 2.1\arcsec~$\times$~1.4\arcsec;
\item Total numbers of observed sky background photoelectrons equal to
      $4.46 \times 10^4$, $6.63 \times 10^4$, and $1.07\times 10^5$ within the elliptical spot during a 30~s visit;
\item 747, 1110, and 1701 photoelectrons per 0.2\arcsec~$\times$~0.2\arcsec~pixel during a 30~s visit; and
\item Standard deviations due to the sky background of approximately $\sqrt{747} = 27.3$, $\sqrt{1110} = 33.3$, and $\sqrt{1701} = 41.2$~photoelectrons per pixel per visit.
\end{itemize}
Within the same 30~s visit time interval, the expected standard deviation in each pixel, in each filter,
due to instrumental background noise is 12.7~photoelectrons~\citep{Jones17}.

\begin{figure*}
\begin{center}
\vspace*{8mm}
\includegraphics[scale=.65]{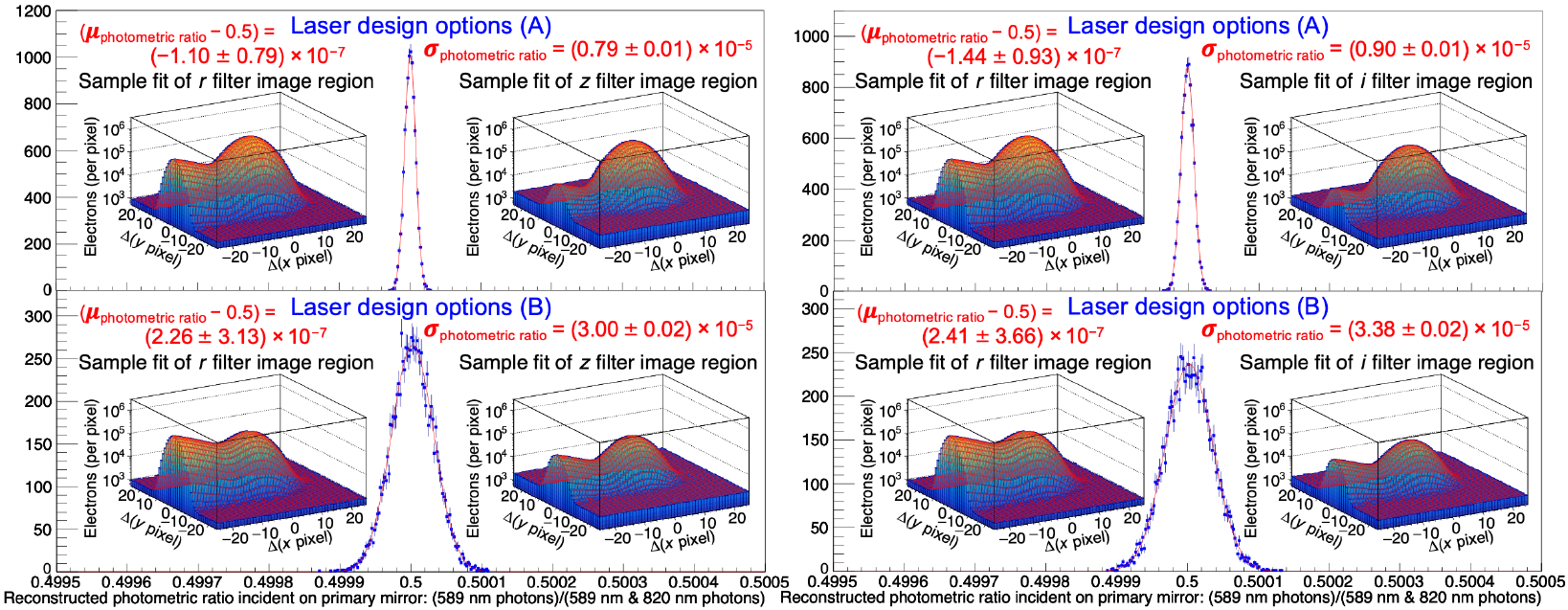}
\end{center}
\caption{Within each of the four sets of plots above, the main plot (i.e., the 1-dimensional fitted Gaussian curve in the centre of each set of plots)
shows the distribution of $10\,000$ fitted photometric ratios, with each ratio reconstructed from pairs of simulated $r$ and $z$ filter image regions
(within the two sets of plots on the left-hand side above), and pairs of simulated $r$ and $i$ filter image regions (within the two sets of plots on the right-hand side above).
Within both the left-hand and right-hand sets of plots, the upper set of plots shows the results of fits with image regions simulated when assuming that the LPRS
is generated with laser design options~(A), with specifications detailed in Section~\ref{sec:Lasers}; and the lower set of plots shows the results when
assuming that the LPRS is instead generated with laser design options~(B).  Within each of the four sets of plots,
each one of the respective $2 \times 10\,000$ simulated image regions consists of a 50~pixel~$\times$~50~pixel square centered around the observed LPRS centroid.  
The inset plots show single examples of simulated $r$ and $z$ image regions (within the two sets of plots on the left),
and simulated $r$ and $i$ image regions (within the two sets of plots on the right), with their respective fits to 2-dimensional Gaussian
LPRS signal ellipses, plus the product of a falling exponential along the $x$-axis and a Gaussian distribution along the $y$-axis
to parametrize the tail of the trail of Rayleigh-scattered background laser light that ``leads up'' to the LPRS spot
(which is perhaps reminiscent of an ``entrance to an igloo'' in all of the inset plots), plus flat background distributions.
These inset plots all have a logarithmic $z$-axis, and linear $x$- and $y$-axes.  The simulated signal photoelectrons,
Rayleigh background photoelectrons, and sky~$+$~noise background photoelectrons, in each simulated pixel within each region,
are generated according to the statistical distributions expected from a single 30~s
LSST visit consisting of a pair of 15~s exposures.  The fitted number of signal photoelectrons is extracted from each image region fit,   
and divided by the (photoelectron)/(incident photon) efficiency (consisting of the expected LSST detector quantum efficiency, multiplied by the expected
throughput fraction of telescope, camera, and filter optics, at 589~nm and at 820~nm for the simulated $r$, $z$, and $i$ filter image regions respectively),
to determine the reconstructed numbers of 589~nm and 820~nm photons incident on the Rubin Observatory primary mirror during the 30~s visit.
For each simulated image region pair, the resulting ratio of reconstructed (589~nm photons)/(589~nm photons + 820~nm photons) is plotted,
and the resulting distribution is fitted to a single Gaussian.  The standard deviations of the fits shown in the main plots within the two \textit{left} sets of plots above, 
which correspond to the expected LPRS photometric ratio statistical uncertainties from a single pair of visits with the $r$ and $z$ filter, are equal to
$(0.79 \pm 0.01) \times 10^{-5}$ in the case of an LPRS that is generated according to laser design options~(A), and to
$(3.00 \pm 0.02) \times 10^{-5}$ in the case of an LPRS that is generated according to laser design options~(B).  
The standard deviations of the fits shown in the main plots within the two \textit{right} sets of plots above, 
which correspond to the expected LPRS photometric ratio statistical uncertainties from a single pair of visits with the $r$ and $i$ filter, are equal to
$(0.90 \pm 0.01) \times 10^{-5}$ in the case of an LPRS that is generated according to laser design options~(A), and to
$(3.38 \pm 0.02) \times 10^{-5}$ in the case of an LPRS that is generated according to laser design options~(B).
The means of the fits that are shown in the main plots within all four of these sets of plots are all
consistent with 0.5.}
\label{fig:photratiosimfit}
\end{figure*}

\section{Resulting Estimated Photometric Ratio Precision}
\label{sec:ResPrec}

Figure~\ref{fig:photratiosimfit} shows the results of sets of numerical simulations to determine the precision of photometric
ratio measurement when using single pairs of 30~s LSST visits to the LPRS spot.  As shown in Fig.~\ref{fig:photratiosimfit},
for pairs of visits in the $r$ and $z$ filters, the resulting estimates of the photometric ratio measurement and its statistical uncertainty
are $0.5 \pm (0.79 \times 10^{-5})$ for the case of an LPRS generated using laser design options~(A),
and $0.5 \pm (3.00 \times 10^{-5})$ for the case of an LPRS generated using laser design options~(B);
and for pairs of visits in the $r$ and $i$ filters, the resulting estimates of the photometric ratio measurement and its statistical uncertainty
are $0.5 \pm (0.90 \times 10^{-5})$ for the case of an LPRS generated using laser design options~(A),
and $0.5 \pm (3.38 \times 10^{-5})$ for the case of an LPRS generated using laser design options~(B).

This shows that in the case of either laser design options~(A) or~(B), one can reach the expected systematic uncertainty limit on the  
LPRS photometric ratio [which is estimated as $\pm(9.0 \times 10^{-5})$,
as stated at the end of Section~\ref{sec:EstFlux}] by utilizing the LPRS together with just a single pair of 30~s LSST visits
either in the $r$ and $z$ filters, or in the $r$ and $i$ filters.

\section{Estimated Impact on Measurements of Dark Energy from Type Ia Supernovae}
\label{sec:EstImp}

We estimate the impact on the precision of upcoming measurements of the dark energy equation of state as a function of redshift, $w(z)$,
from the photometric ratio calibration provided by this LPRS.  Our analysis proceeds in an analogous way to the impact analysis that is
described in Section~\href{https://arxiv.org/pdf/2001.10958.pdf#sec9}{9} of \citetalias{Alb20a}.  In particular, we also use the typical parametrization
$w(z) = w_0 + \frac{z}{1 + z}w_a$ here (where the quantities $w_0$ and $w_a$ respectively parameterize the equation of state of dark
energy at the present time, and the amount of change in the equation of state of dark energy over cosmic history).  And again we use
the typical figure of merit $\mathcal{F}_{\rm DE} \equiv [{\rm det}\,\mathbf{C}(w_0,w_a)]^{-\frac{1}{2}}$, where $\mathbf{C}(w_0,w_a)$ is
the covariance matrix of the $(w_0,w_a)$ estimations, to characterise the expected performance of a measurement of the properties of
dark energy.

We additionally use a nearly identical procedure to that described in Section~\href{https://arxiv.org/pdf/2001.10958.pdf#sec9}{9} of
\citetalias{Alb20a}, in order
to generate the simulated dataset catalog of SNeIa which represents the expected LSST observations with the photometric calibration
that would result from the LPRS described in this present paper.  The sole differences in the catalog generation here are from the
larger expected improvements in the SNeIa apparent magnitude uncertainties that would result from the two-laser LPRS that is described
in this paper:
\begin{itemize}[label = \textbullet, leftmargin = 4mm, labelsep = *, itemsep = 1mm, topsep = 1mm]
\item When generating the simulated SNeIa catalog that represents expected LSST observations \underline{\textbf{with}} LPRS-based
      photometric calibration, the generated systematic uncertainties on the SNeIa magnitudes are reduced by a factor of 4.01 from
      those in the joint light-curve analysis (JLA) that is described in~\citet{Betoule14},
      corresponding to the expected improvement in SNeIa magnitude measurement from photometric calibration from this LPRS;
\item Also when generating the simulated SNeIa catalog that represents expected LSST observations \underline{\textbf{with}} photometric
      calibration from this LPRS, the generated systematic covariances between the SNeIa magnitude and light-curve stretch values, as
      well as between the SNeIa magnitude and colour parameter values, are similarly reduced by a factor of 2.00 from those in the JLA,  
      corresponding to the expected improvement in SNeIa magnitude measurement from this LPRS-based photometric calibration.
\end{itemize}
An identical catalog to that used in the analysis described in Section~\href{https://arxiv.org/pdf/2001.10958.pdf#sec9}{9} of 
\citetalias{Alb20a} is used to   
represent expected LSST observations \underline{\textbf{without}} LPRS-based photometric calibration.  And, an identical fitting
strategy to the one that is described in that Section is used here for the fits to both simulated SNeIa catalogs.

\begin{figure}
\begin{center}
\includegraphics[scale=0.52]{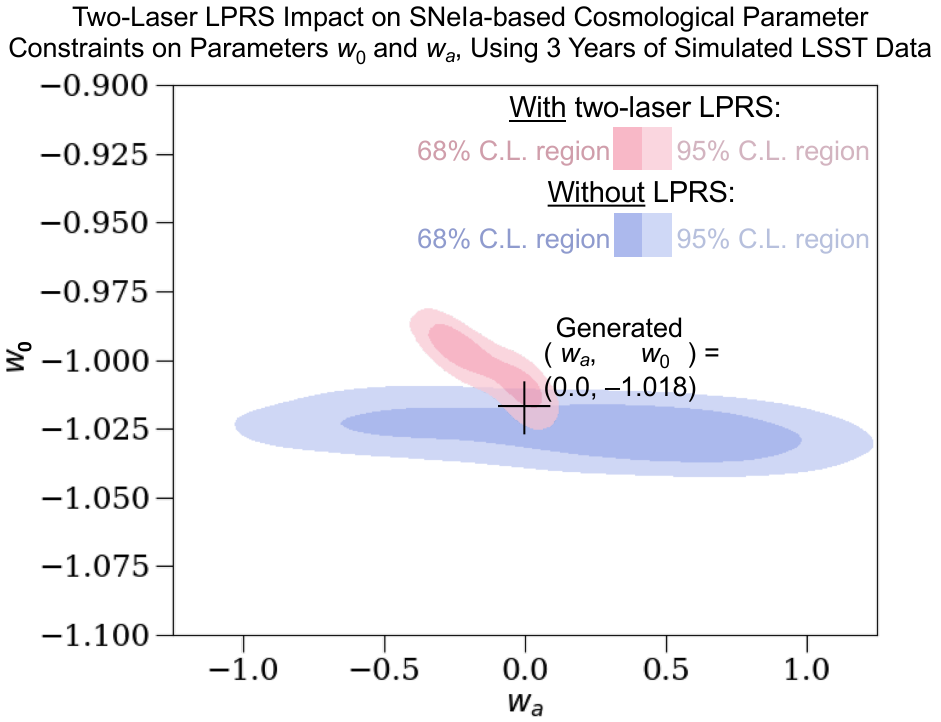}
\end{center}
\caption{Constraints on the dark energy equation of state parameters $w_0$ and $w_a$, obtained using
simulated catalogues of type Ia supernovae.  Each one of the two SNeIa catalogues that are fitted to obtain
these constraints contains $120\,000$ simulated SNeIa (corresponding to approximately 3 years of observation at the Rubin Observatory).
The generation of the two simulated SNeIa catalogs, as well as the fits that are performed to each catalog, are
implemented using the {\tt CosmoSIS} cosmological parameter estimation code~\citep{CosmoSIS}, and are explained
in the text here, as well as in Section~\href{https://arxiv.org/pdf/2001.10958.pdf\#sec9}{9} of \citetalias{Alb20a}.
The resulting values of the figure of merit
parameter $\mathcal{F}_{\rm DE}$ for the fits are 313 for the fit to the SNeIa catalog representing expected LSST   
observations without LPRS-based photometric calibration, and 1646 for the fit to the SNeIa catalog representing
expected LSST observations with photometric calibration from the two-laser LPRS described in this paper.
(Also, as would be expected on average, the central values of each fit are approximately one standard
deviation away from the generated values of $w_0$ and $w_a$.)
The fits represent a $\frac{1646}{313} = 5.26$-fold
expected improvement in the dark energy figure of merit parameter $\mathcal{F}_{\rm DE}$ from LPRS-based photometric
calibration (with even greater expected
resulting $\mathcal{F}_{\rm DE}$ increases for SNeIa datasets that correspond to greater than 3 years of Rubin Observatory
observation).
}
\label{fig:LPRS2PhotImpactOnDECosmologyPlot}
\end{figure}

The results of the fits to the two simulated catalogs, when projected onto the $(w_0,w_a)$ plane (and, thus, when
marginalised over all of the other fitted parameters), are shown in Figure~\ref{fig:LPRS2PhotImpactOnDECosmologyPlot}.
The resulting values of the figure of merit
parameter $\mathcal{F}_{\rm DE}$ for the fits are 313 for the fit to the SNeIa catalog representing expected LSST  
observations without LPRS-based photometric calibration, and 1646 for the fit to the SNeIa catalog representing
expected LSST observations with LPRS-based photometric calibration, representing a $\frac{1646}{313} = 5.26$-fold
expected improvement in the dark energy figure of merit parameter $\mathcal{F}_{\rm DE}$ from photometric
calibration from this LPRS over 3 years of Rubin Observatory observation.  (Even larger resulting $\mathcal{F}_{\rm DE}$ increases
due to this LPRS-based photometric calibration would be expected for SNeIa datasets that correspond to greater than
3 years of Rubin Observatory observation.)

\section{Conclusions, and Comparison of LPRS Techniques}

In this paper, together with \citetalias{Alb20a}, we present two
methods for establishing a reference for relative photometry between the visible and NIR
(and specifically between photometry at 589/590~nm and 819/820~nm wavelengths) of unprecedented precision
using mountaintop-located laser sources to excite neutral atoms of sodium in the mesospheric sodium layer.
The method that is described in \citetalias{Alb20a} would utilise a single laser tuned to the 342.78~nm excitation wavelength
of neutral atomic sodium, and would require this laser to have very high ($\ge$500~W) optical output power; whereas the method that
is described in this paper would utilise two lasers, respectively tuned to 3.9~GHz below and above the neutral sodium resonances
at 589.16~nm and 819.71~nm wavelengths, and would permit much lower laser optical output power (within a 10~--~30~W output power
range that is available with some present off-the-shelf dye laser systems).  The method that is described in this paper would,
however, require new polarization filters to be installed into the telescope camera in order to sufficiently remove laser atmospheric
Rayleigh backscatter from the resulting telescope images.

As we have shown in this paper, when implemented this method described here will improve measurements
of dark energy from type~Ia supernovae, using upcoming surveys such as
the first 3 years of observations at the Vera C.~Rubin Observatory, by
approximately a factor of $5.26$ for the standard dark energy ``figure of merit'' $\mathcal{F}_{\rm DE}$ (which is based on the expected
uncertainties on measurements of the dark energy equation of state parameters $w_0$ and $w_a$).
The LPRS technique that is described in this paper, when compared with the technique described in \citetalias{Alb20a},
would provide a far greater improvement in the measurements of these dark energy parameters, due to the fact
that the resulting LPRS would be over a factor of $10^3$ brighter than the LPRS described in \citetalias{Alb20a}, thus in effect
removing limitations from observed LPRS photon statistics.
(And also, we note, could be continuously dimmed down across that full range of brightness, for use in testing of the linearity of the
relative photometry.)  Additionally, the two-laser LPRS described in this paper would be far less challenging and expensive to
construct than the one-laser LPRS that is described in \citetalias{Alb20a}, due to the much
lower required \mbox{optical} output power of the two lasers, when compared with the output power required of the single-laser LPRS.
Thus, we prefer and recommend the development and testing of the two-laser LPRS that is described in this paper.
(We must also note that if time and cost were no issue, one would, of course, prefer the development and testing of both options,   
especially since the one-laser LPRS does have an interesting comparative advantage of not requiring polarization filters to be
installed in the telescope camera.  However, given the choice, the two-laser LPRS would be, by far, the simpler and more effective
option of the two.)

The two-laser LPRS that is described in this paper uses the STIRAP (STImulated Raman Adiabatic Passage) process to
excite mesospheric neutral sodium atoms.
STIRAP is a well-established technique developed 30 years ago, and demonstrated in thousands of results in
laboratories across the world that have been documented in over 200 publications and multiple review articles since 1990
[as examples,~\citet{Ber15}, and~\citet{Vit17}];
however the STIRAP technique has not yet been demonstrated in the open atmosphere.  Thus, this will be a
novel challenge; however, when complete, the implementation of this two-laser LPRS may thus mark the
first utilization and observation of ``STIRAP in the sky'' --- an important milestone in the progress of the
STIRAP technique, in addition to the specific usage of this LPRS for calibration of unprecedented precision
in cosmology, astronomy, and atmospheric physics.

\section*{Acknowledgements}

The authors would like to thank Prof.~Gabriele Ferrari of Universit{\`a} di Trento and of LEOSolutions (Rovereto, Italy)
for critical and useful discussions regarding parametric crystal options for laser wavelength tunability. 
We would also like to thank Dr.~Andrew MacRae of the
University of Victoria for reading over the manuscript and providing extremely helpful comments and suggestions,
and Prof.~Christopher Stubbs of Harvard University for his helpful encouragement at an early stage of these papers.
JEA gratefully acknowledges support from Canadian Space Agency grants 19FAVICA28 and 17CCPVIC19.

\section*{Data Availability}

All code and data generated and used for the results of this paper is available from the authors upon request.
 



\bibliographystyle{mnras}
\bibliography{lidasp2} 

\begin{thebibliography}{}
\makeatletter
\relax
\def\mn@urlcharsother{\let\do\@makeother \do\$\do\&\do\#\do\^\do\_\do\%\do\~}
\def\mn@doi{\begingroup\mn@urlcharsother \@ifnextchar [ {\mn@doi@}
  {\mn@doi@[]}}
\def\mn@doi@[#1]#2{\def\@tempa{#1}\ifx\@tempa\@empty \href
  {http://dx.doi.org/#2} {doi:#2}\else \href {http://dx.doi.org/#2} {#1}\fi
  \endgroup}
\def\mn@eprint#1#2{\mn@eprint@#1:#2::\@nil}
\def\mn@eprint@arXiv#1{\href {http://arxiv.org/abs/#1} {{\tt arXiv:#1}}}
\def\mn@eprint@dblp#1{\href {http://dblp.uni-trier.de/rec/bibtex/#1.xml}
  {dblp:#1}}
\def\mn@eprint@#1:#2:#3:#4\@nil{\def\@tempa {#1}\def\@tempb {#2}\def\@tempc
  {#3}\ifx \@tempc \@empty \let \@tempc \@tempb \let \@tempb \@tempa \fi \ifx
  \@tempb \@empty \def\@tempb {arXiv}\fi \@ifundefined
  {mn@eprint@\@tempb}{\@tempb:\@tempc}{\expandafter \expandafter \csname
  mn@eprint@\@tempb\endcsname \expandafter{\@tempc}}}

\bibitem[\protect\citeauthoryear{Albert et~al.}{Albert et~al.}{2021}]{Alb20a}
Albert J.~E.,  et~al., 2021,
  {\MYhref[magenta]{https://doi.org/10.1093/mnras/stab1621}{\textbf{MNRAS}}}
  \textbf{508}, {{4399
  (\MYhref[blue]{https://arxiv.org/abs/2001.10958}{\tt{arXiv:2001.10958}};
  \unskip\parfillskip 0pt \par }
  \MYhref[blue]{https://doi.org/10.1093/mnras/stab1621}{Paper~I}).}

\bibitem[\protect\citeauthoryear{Amplitude}{Amplitude}{2021}]{Amplitude}
Amplitude {Laser Group~(Pessac, France), 2021}, {product Powerlite DLS 9050
  laser,
  \url{https://amplitude-laser.com/wp-content/uploads/2019/02/Powerlite-DLS-90%
00_ref-f_BD.pdf}.}

\bibitem[\protect\citeauthoryear{Bergmann, Vitanov  \& Shore}{Bergmann
  et~al.}{2015}]{Ber15}
Bergmann K.,  Vitanov N.~V.,   Shore B.~W.,  2015,
  {\MYhref[magenta]{https://doi.org/10.1063/1.4916903}{\textbf{J. Chem.
  Phys.}}} \textbf{142}, 170901.

\bibitem[\protect\citeauthoryear{Betoule et~al.}{Betoule
  et~al.}{2014}]{Betoule14}
Betoule M.,  et~al., 2014,
  {\MYhref[magenta]{https://doi.org/10.1051/0004-6361/201423413}{\textbf{A\&A}%
}} \textbf{568}, A22.

\bibitem[\protect\citeauthoryear{Budker, Kimball  \& DeMille}{Budker
  et~al.}{2008}]{Bud08}
Budker D.,  Kimball D.~F.,   DeMille D.~P.,  2008, Atomic Physics: An
  Exploration Through Problems and Solutions, 2\textsuperscript{nd} edn.
Oxford Univ. Press, Oxford, UK (ISBN
  \MYhref[blue]{https://isbndb.com/book/9780199532414}{978-0199532414}).

\bibitem[\protect\citeauthoryear{Calia, Hackenberg, Holzl{\"o}hner, Lewis  \&
  Pfrommer}{Calia et~al.}{2014}]{Cal14}
Calia D.~B.,  Hackenberg W.,  Holzl{\"o}hner R.,  Lewis S.,   Pfrommer T.,
  2014, {\MYhref[magenta]{https://doi.org/10.1515/aot-2014-0025}{\textbf{Adv.
  Opt. Technol.}}} \textbf{3}, 345.

\bibitem[\protect\citeauthoryear{Connor et~al.}{Connor et~al.}{2017}]{Connor17}
Connor T.,  et~al., 2017,
  {\MYhref[magenta]{https://doi.org/10.3847/1538-4357/aa8ad5}{\textbf{ApJ}}}
  \textbf{848}, 37.

\bibitem[\protect\citeauthoryear{Danileiko, Romanenko  \& Yatsenko}{Danileiko
  et~al.}{1994}]{Dan94}
Danileiko M.~V.,  Romanenko V.~I.,   Yatsenko L.~P.,  1994,
  {\MYhref[magenta]{https://doi.org/10.1016/0030-4018(94)90499-5}{\textbf{Opt.
  Commun.}}} \textbf{109}, 462.

\bibitem[\protect\citeauthoryear{EdgeWave}{EdgeWave}{2021}]{EdgeWave}
EdgeWave {GmbH~(W{\"u}rselen, Germany), 2021}, {product nos.~ISxxx-2 green
  IS-series lasers,
  \url{https://www.edge-wave.de/web/wp-content/uploads/ISweb.pdf}.}

\bibitem[\protect\citeauthoryear{Fan, Zhou  \& Feng}{Fan et~al.}{2016}]{Fan16}
Fan T.,  Zhou T.,   Feng Y.,  2016,
  {\MYhref[magenta]{https://doi.org/10.1038/srep19859}{\textbf{Sci. Rep.}}}
  \textbf{6}, 19859.

\bibitem[\protect\citeauthoryear{Gaubatz, Rudecki, Schiemann  \&
  Bergmann}{Gaubatz et~al.}{1990}]{Gau90}
Gaubatz U.,  Rudecki P.,  Schiemann S.,   Bergmann K.,  1990,
  {\MYhref[magenta]{https://doi.org/10.1063/1.458514}{\textbf{J. Chem. Phys.}}}
  \textbf{92}, 5363.

\bibitem[\protect\citeauthoryear{H{\"o}hn}{H{\"o}hn}{1969}]{Hohn69}
H{\"o}hn D.~H.,  1969,
  {\MYhref[magenta]{https://doi.org/10.1364/AO.8.000367}{\textbf{Appl. Opt.}}}
  \textbf{8}, 367.

\bibitem[\protect\citeauthoryear{Johnson, Allen, Hicks  \& Burdin}{Johnson
  et~al.}{2010}]{Joh10}
Johnson J.~B.,  Allen S.~D.,  Hicks J.~L.,   Burdin J.,  2010, in
  \MYhref[blue]{https://doi.org/10.1117/12.850263}{Chemical, Biological,
  Radiological, Nuclear, and Explosives (CBRNE) Sensing XI},
  \MYhref[magenta]{https://doi.org/10.1117/12.850263}{\textbf{Proc.~SPIE}}
  \textbf{7665}, p. 766512.

\bibitem[\protect\citeauthoryear{Jones}{Jones}{2017}]{Jones17}
Jones L.,  2017, Technical report, ``Calculating LSST Limiting Magnitudes and
  SNR''.
LSST / Rubin Observatory Simulation Technical Note 002 (SMTN-002). Available
  at: \url{https://smtn-002.lsst.io}.

\bibitem[\protect\citeauthoryear{Jones et~al.}{Jones et~al.}{2018}]{Jones18}
Jones D.~O.,  et~al., 2018,
  {\MYhref[magenta]{https://doi.org/10.3847/1538-4357/aab6b1}{\textbf{ApJ}}}
  \textbf{857}, 51.

\bibitem[\protect\citeauthoryear{Jones et~al.}{Jones et~al.}{2019}]{Jones19}
Jones L.,  et~al., 2019, Technical report, ``Systems Engineering-approved LSST
  Throughputs Repository''.
Available at: \url{https://github.com/lsst-pst/syseng\_throughputs}.

\bibitem[\protect\citeauthoryear{Kane, Hillman  \& Denman}{Kane
  et~al.}{2014}]{Kan14}
Kane T.~J.,  Hillman P.~D.,   Denman C.~A.,  2014, in
  \MYhref[blue]{https://doi.org/10.1117/12.2055632}{Adaptive Optics Systems
  IV},
  \MYhref[magenta]{https://doi.org/10.1117/12.2055632}{\textbf{Proc.~SPIE}}
  \textbf{9148}, p. 91483G.

\bibitem[\protect\citeauthoryear{Kelleher \& Podobedova}{Kelleher \&
  Podobedova}{2008}]{Kel08}
Kelleher D.~E.,  Podobedova L.~I.,  2008,
  {\MYhref[magenta]{https://doi.org/10.1063/1.2735328}{\textbf{J. Phys. Chem.
  Ref. Data}}} \textbf{37}, {267. Please note that this reference contains the
  correct vacuum wavelengths for sodium transitions. In reference
  \citet{NISTtables} below, when a wavelength in vacuum for a sodium transition
  is provided, it is often actually the wavelength in air at standard pressure
  instead.}

\bibitem[\protect\citeauthoryear{Kirk et~al.}{Kirk et~al.}{2015}]{Kirk15}
Kirk D.,  et~al., 2015,
  {\MYhref[magenta]{https://doi.org/10.1093/mnras/stv1268}{\textbf{MNRAS}}}
  \textbf{451}, 4424.

\bibitem[\protect\citeauthoryear{Kramida, Ralchenko, Reader  et~al.}{Kramida
  et~al.}{2020}]{NISTtables}
Kramida A.,  Ralchenko Y.,  Reader J.,   et~al., 2020, Technical report, Atomic
  Spectra Database (ver. 5.8), NIST.
Available at: \url{https://physics.nist.gov/asd}. [Please, however, see the
  note in the reference for \citet{Kel08} above.]

\bibitem[\protect\citeauthoryear{Pedreros~Bustos et~al.}{Pedreros~Bustos
  et~al.}{2018}]{Ped18}
Pedreros~Bustos F.,  et~al., 2018,
  {\MYhref[magenta]{https://doi.org/10.1038/s41467-018-06396-7}{\textbf{Nat.
  Commun.}}} \textbf{9}, 3981.

\bibitem[\protect\citeauthoryear{Pedreros~Bustos et~al.}{Pedreros~Bustos
  et~al.}{2020}]{Ped20}
Pedreros~Bustos F.,  et~al., 2020,
  {\MYhref[magenta]{https://doi.org/10.1364/JOSAB.389007}{\textbf{J. Opt. Soc.
  Am. B}}} \textbf{37}, 1208.

\bibitem[\protect\citeauthoryear{{Radiant Dyes}}{{Radiant
  Dyes}}{2021}]{RadiantDyes}
{Radiant Dyes} {Laser GmbH~(Wermelskirchen, Germany), 2021}, {product
  NarrowScan High Rep.~Laser,
  \url{https://www.radiant-dyes.com/PDF/6\_NarrowScan\%20high\%20Rep\_Juni\%20%
2011.pdf}.}

\bibitem[\protect\citeauthoryear{Rochester}{Rochester}{2021}]{ADM}
Rochester S.~M.,  2021, The {\tt Atomic Density Matrix} software package can be
  downloaded from \url{http://rochesterscientific.com/ADM/}; and was initially
  described in Rochester, S. M.
  ``\MYhref[blue]{https://escholarship.org/uc/item/6cg0m520}{Modeling Nonlinear
  Magneto-optical Effects in Atomic Vapors}'' (Ph.D. dissertation, 2010;
  University of California, Berkeley, CA, USA).

\bibitem[\protect\citeauthoryear{{Sirah}}{{Sirah}}{2021}]{Sirah}
{Sirah} {Lasertechnik GmbH~(Grevenbroich, Germany), 2021}, {product Double Dye
  Laser, \url{http://www.sirah.com/wp-content/uploads/documents/DoubleDye.pdf};
  and product Pulsed Amplifier 2xHRR,
  \url{http://www.sirah.com/wp-content/uploads/2021/03/PulsedAmp-2xHRR.pdf}.}

\bibitem[\protect\citeauthoryear{{Thorlabs}}{{Thorlabs}}{2021}]{Thorlabs}
{Thorlabs} {Inc.~(Newton, NJ, USA), 2021}, {product no.~CCM1-PBS25-532-HP/M,
  \url{https://www.thorlabs.com/thorproduct.cfm?partnumber=CCM1-PBS25-532-HP/M%
}; and product no.~DMLP650L,
  \url{https://www.thorlabs.com/thorproduct.cfm?partnumber=DMLP650L}.}

\bibitem[\protect\citeauthoryear{Vitanov, Rangelov, Shore  \& Bergmann}{Vitanov
  et~al.}{2017}]{Vit17}
Vitanov N.~V.,  Rangelov A.~A.,  Shore B.~W.,   Bergmann K.,  2017,
  {\MYhref[magenta]{https://doi.org/10.1103/RevModPhys.89.015006}{\textbf{Rev.
  Mod. Phys.}}} \textbf{89}, 015006.

\bibitem[\protect\citeauthoryear{Vogt et~al.}{Vogt et~al.}{2017}]{Vog17}
Vogt F. P.~A.,  et~al., 2017,
  \MYhref[magenta]{https://doi.org/10.1103/PhysRevX.7.021044}{\textbf{Phys.
  Rev. X}} \textbf{7}, 021044.

\bibitem[\protect\citeauthoryear{Wood-Vasey et~al.}{Wood-Vasey
  et~al.}{2007}]{WoodVasey07}
Wood-Vasey W.~M.,  et~al., 2007,
  {\MYhref[magenta]{https://doi.org/10.1086/518642}{\textbf{ApJ}}}
  \textbf{666}, 694.

\bibitem[\protect\citeauthoryear{Zuntz et~al.}{Zuntz et~al.}{2015}]{CosmoSIS}
Zuntz J.,  et~al., 2015,
  {\MYhref[magenta]{https://doi.org/10.1016/j.ascom.2015.05.005}{\textbf{Astro%
n. Comput.}}} \textbf{12}, 45 (with code repository and instructions available
  at \url{https://bitbucket.org/joezuntz/cosmosis/wiki/Home}).

\makeatother
\end{thebibliography}





\bsp	
\label{lastpage}
\end{document}